\documentclass[pra,floatfix,twocolumn,amsfonts,amsmath,amssymb,nofootinbib,reprint,superscriptaddress]{revtex4-2}
\pdfoutput=1

\usepackage{graphicx}
\usepackage{dcolumn}
\usepackage{bm}
\usepackage[utf8]{inputenc}

\usepackage[colorlinks]{hyperref}
\usepackage[normalem]{ulem}
\usepackage{comment}
\usepackage{xcolor}
\usepackage{amsthm}
\usepackage{mathtools}
\usepackage{bbm}
\usepackage{bm}
\usepackage{graphicx}
\usepackage{tikz}
\usetikzlibrary{arrows}
\usepackage{soul}

\newtheorem{theorem}{Theorem}

\newtheorem{definition}[theorem]{Definition}

\def\bra#1{\mathinner{\langle{#1}|}}
\def\ket#1{\mathinner{|{#1}\rangle}}
\def\braket#1#2{\mathinner{\langle{#1|#2}\rangle}}
\def\ketbra#1#2{\mathinner{|{#1}\rangle\!\langle{#2}|}}

\def\BraVert{\egroup\,\mid@vertical\,\bgroup}

\def\dket#1{\mathinner{|{#1}\rangle\!\rangle}}
\def\dketbra#1#2{\mathinner{|{#1}\rangle\!\rangle\!\langle\!\langle{#2}|}}

\DeclareMathOperator{\Tr}{Tr}

\renewcommand\L{\mathcal{L}}

\newcommand{\M}{\mathcal{M}}
\newcommand{\HS}{\mathcal{H}}

\newcommand{\id}{\mathbbm{1}}

\usepackage[normalem]{ulem}

\definecolor{darkred}{rgb}{0.78, 0.03, 0.08}

\begin{document}

\preprint{APS/123-QED}

\title{Simulating Noncausality with Quantum Control of Causal Orders}

\author{Anna Steffinlongo}
\affiliation{ICFO-Institut de Ciencies Fotoniques, The Barcelona Institute of Science and Technology,\\ 08860 Castelldefels, Barcelona, Spain}
\author{Hippolyte Dourdent}
 \email{hippolyte.dourdent@icfo.eu}
\affiliation{ICFO-Institut de Ciencies Fotoniques, The Barcelona Institute of Science and Technology,\\ 08860 Castelldefels, Barcelona, Spain}

\date{\today}

\begin{abstract}
Logical consistency with free local operations is compatible with non-trivial classical communications, where all parties can be both in each other's past and future—a phenomenon known as noncausality. Noncausal processes, such as the ``Lugano (AF/BW) process", violate causal inequalities, yet their physical realizability remains an open question. In contrast, the quantum switch—a physically realizable process with indefinite causal order—can only generate causal correlations. Building on a recently established correspondence  [Kunjwal \& Baumeler, PRL 131, 120201 (2023)] between the SHIFT measurement, which exhibits nonlocality without entanglement, and the Lugano process, we demonstrate that the SHIFT measurement can be implemented using a quantum switch of classical communications in a scenario with quantum inputs. This shows that the structure of the Lugano process can be simulated by a quantum switch and that successful SHIFT discrimination witnesses causal nonseparability rather than noncausality. Finally, we identify a broad class of ``superposition of classical communications" derived from classical processes without global past capable of realizing similar causally indefinite measurements. We examine these results in relation to the ongoing debate on implementations of indefinite causal orders.
\end{abstract}

\maketitle

\medskip

\section{Introduction}

\subsection{State of the art}
In local operations and classical communication (LOCC) protocols, each party can perform arbitrary quantum operations on their system, while communication remains classical \cite{chitambar14} with a causally well-defined order on the operations. In single-round scenarii, LOCC can enable specific joint measurements. For instance, the basis $\{\ket{00},\ket{01},\ket{1+},\ket{1-}\}$ cannot be measured locally by Alice and Bob using only shared randomness \cite{chitambar14,pauwels24}. However, it can be with LOCC: Alice measures her qubit in the $\{\ket{0},\ket{1}\}$ basis, sends her outcome $a$ to Bob, who then measures in $\{\ket{0},\ket{1}\}$ if $a=0$ or in  $\{\ket{\pm}=(\ket{0}\pm\ket{1})/\sqrt{2}\}$ if $a=1$.

Nevertheless, certain product-state bases remain locally unmeasurable under LOCC \cite{bennett_99}, a phenomenon termed \textit{``nonlocality without entanglement''} (NLWE). The \textit{SHIFT basis} exemplifies such an ensemble:
\begin{align}
    &\{\ket{000},\ket{+01},\ket{01+},\ket{01-},\notag\\ 
&\ket{1+0},\ket{-01},\ket{1-0},\ket{111}\}
\label{eq:shift}
\end{align}

\textit{Classical communications without definite causal orders} can bypass NLWE. In bipartite scenarii, local operations within classical causal loops (LOCC*) coincides with separable operations \cite{akibue17}. More generally, operational closed-timelike curves based on post-selected teleportations \cite{politzer,bennett,svetlichny} enable perfect discrimination of any set of linearly independent quantum states \cite{brun11}. However, these structures are often viewed as pathological due to their ability to  solve complex problems efficiently \cite{brun11,araujo3} and generate logical antinomies under free operations \cite{baumeler, baumeler21}. 

To avoid these pathologies, the ``process matrix formalism'' \cite{oreshkov1} was developed to describe global processes while preserving the validity of ordinary local physics laws. Its classical, deterministic counterpart—``\textit{process functions}'' \cite{baumeler16,baumeler19} models consistent classical communication under free local operations. A simple characterization exists for all $N-$partite processes \cite{tobar} and, remarkably, process functions without definite causal order have been identified for $N\geq 3$ \cite{baumeler2,af,tobar}. These, termed  \textit{``noncausal''}, generate correlations that violate causal inequalities \cite{oreshkov1,branciard1}.

Kunjwal and Baumeler \cite{kunjwal23a} showed that the SHIFT basis corresponds to a  tripartite noncausal process function -- the \textit{Lugano (AF/BW) process} \cite{af,baumeler2}, where each party's input depends non-trivially on the others' outputs, with no global past, future, or restrictions on their local operations, 
\begin{align}   x:=c(b\oplus 1),\hspace{5mm}  y:=a(c\oplus 1),\hspace{5mm}  z:=b(a\oplus 1) \label{eq:lugano} \end{align} with $x,y,z$ and $a,b,c$ respectively labeling the input and output bits of Alice, Bob, and Charlie. Local operations with the Lugano process implement the projective measurement in the SHIFT basis, and this SHIFT measurement \textit{simulates} the Lugano process by implementing its classical channel in a causal quantum circuit which, combined with retro-causal identity channels, realizes the noncausal process. This advantage of Local Operations with Process Function ``LOPF" over LOCC extends to to a broader class of multipartite noncausal Boolean processes without global past, which can be transformed into NLWE multiqubit bases, trading the causality of correlations for local measurability.

Whether these noncausal structures are physically realizable remains open. The Lugano process is unitarily extendible, a criterion for physical implementability  ~\cite{araujo4}, and can be modeled via time-delocalized subsystems \cite{oreshkov19,Wechs23}, observer-dependent local events \cite{guerin}, or a routed quantum circuit \cite{vanrietvelde23}. 
However, causal inequality violations are unlikely in fixed spacetime obeying relativistic causality \cite{vilasini24,vilasini24a}, suggesting reliance on closed timelike curves. Known physically realizable processes with indefinite causal order -- quantum circuits with quantum control of causal orders (QC-QCs) \cite{wechs1}, including the ``quantum switch'' \cite{chiribella13} -- only produce causal correlations \cite{araujo1,oreshkov16,wechs1,purves21}. Whether the Lugano process emerges in general relativity or at the interface of quantum theory and gravity \cite{hardy05,hardy07}, and how to realize the SHIFT measurement  locally, remains \textit{a priori} unknown. \\

\subsection{Contribution} We demonstrate how the SHIFT measurement can also be implemented by local operations with quantum control of classical communications. 
Equivalently, the noncausal structure of the Lugano process can be \textit{simulated} with such a quantum switch in a scenario with quantum inputs \cite{dourdent21}. This implies that successful SHIFT discrimination is only an ``operational signature of noncausal correlations''~\cite{kunjwal23a} in the specific LOPF scenario.  More generally, implementing the SHIFT measurement rather certifies the weaker notion of causal nonseparability, namely the inability to express a process as a convex mixture of processes with definite causal order. We derive a QC-QC similar to the quantum switch from a unitary extension of the Lugano process matrix, yielding a SHIFT implementation  we term “local operations with superposition of classical communications (LOSupCC).”
We identify a broad class of such ``SupCC" processes that implement measurements exhibiting both NLWE and causal nonseparability, and relate our results to the ongoing debate on implementating indefinite causal orders (see Appendix \ref{app:exp}).
\medskip

\section{Results}

\subsection{The SHIFT-Lugano correspondence}\label{sec:eq}
We present a slightly modified but equivalent formulation of the SHIFT-Lugano correspondence from \cite{kunjwal23a}, focusing on constructing the SHIFT measurement via local operations using the Lugano process. The construction of the  classical channel underlying the Lugano process via implementing a SHIFT measurement in a causal quantum circuit is provided in Appendix \ref{app:lugshift}.
\medskip

\begin{figure}[ht]
	\begin{center}
	\includegraphics[width=0.9\columnwidth]{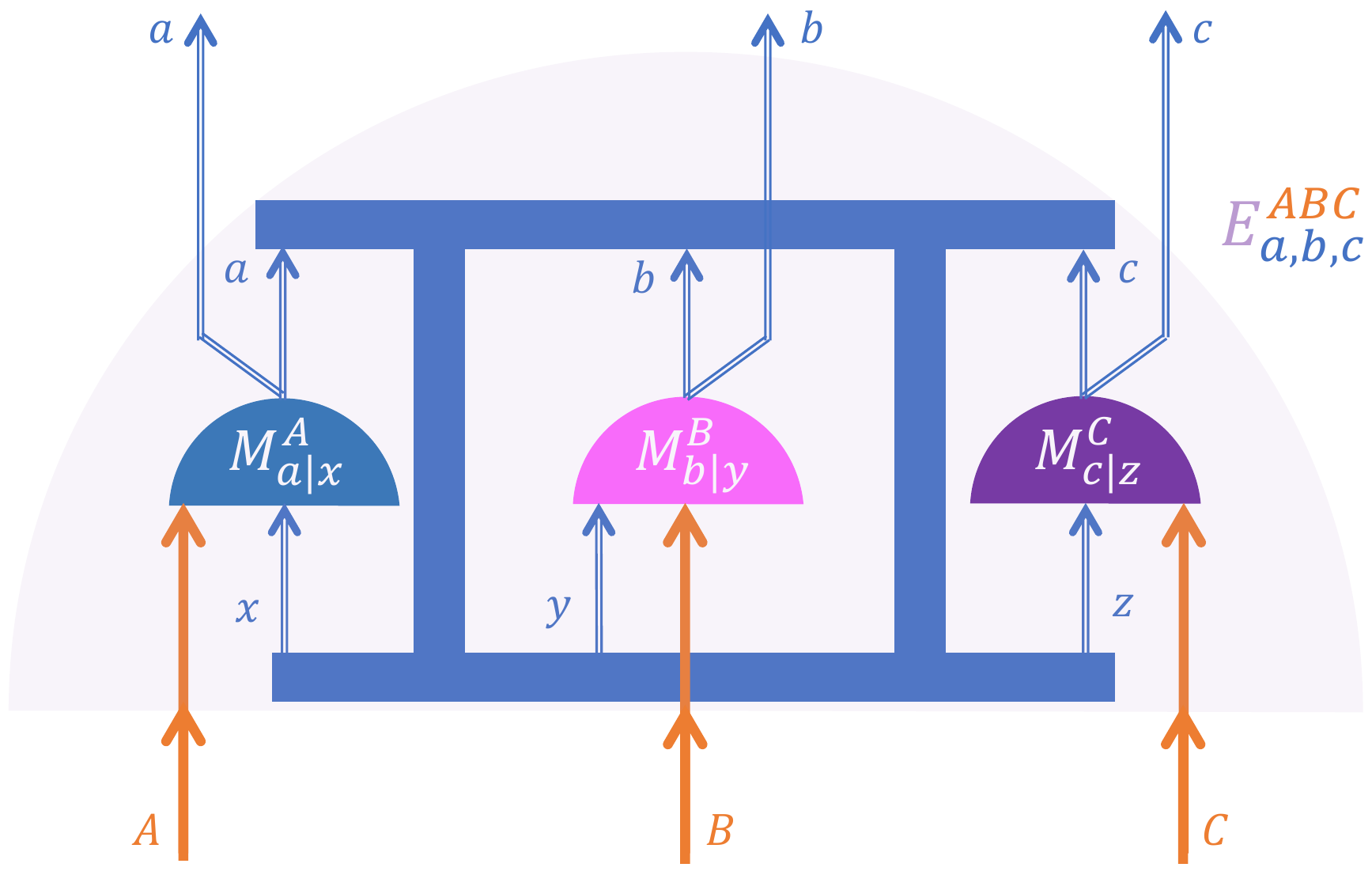}
	\end{center}
	\caption{The LOPF scenario: three parties perform local (projective) operations $(M_{a|x}^A)_a,(M_{b|y}^B)_b,(M_{c|z}^C)_c$ on separated quantum systems (orange wires). The Boolean process function (blue comb) maps the parties' outputs $(a,b,c)$ into their respective inputs $(x,y,z)$, generating an effective projective measurement $(E_{a,b,c}^{ABC})_{a,b,c}$.  } 
	\label{fig:locc}
\end{figure}

Consider the tripartite scenario in Fig.~\ref{fig:locc}. Alice, Bob and Charlie control separate labs with input spaces $\mathcal{H}^A$, $\mathcal{H}^B$, and $\mathcal{H}^C$. Throughout, we restrict to finite-dimensional systems and, in particular, assume that these local Hilbert spaces are two-dimensional, so that each party holds a qubit. Each performs a fixed local quantum projective measurement, denoted  $(M_{a|x}^A=\ketbra{a|x}{a|x}^A)_a,(M_{b|y}^B=\ketbra{b|y}{b|y}^B)_b$ and $(M_{c|z}^C=\ketbra{c|z}{c|z}^C)_c$, with settings $x,y,z\in\{0,1\}$ and outcomes $a,b,c\in\{0,1\}$. (We denote the space of linear operators on $\HS^X$ as $\L(\HS^X)$ and write concisely $\HS^{XY} = \HS^X\otimes\HS^Y$, superscripts indicate on what spaces operators act.) Furthermore, we assume that these local operations are connected via a Boolean \textit{process function} (see \cite{kunjwal23a}, Appendix). In the tripartite case, this process can be expressed as a set of functions mapping the outputs of two parties into the input of the third, $\{ w_A:(b,c)\rightarrow x \hspace{1mm};\hspace{1mm} w_B:(c,a)\rightarrow y \hspace{1mm};\hspace{1mm} w_C: (a,b)\rightarrow z\}$, such that for every fixed output of one party, at most one-way signalling is possible between the other two \cite{baumeler19,tobar}. The Theorem of \cite{kunjwal23a} shows that this configuration yields an effective measurement \cite{supic17,hoban18,dourdent21,dourdent24} $(E_{a,b,c}^{ABC})_{a,b,c}$ 
with rank-1 projectors:
\begin{align}
E_{a,b,c}^{ABC}=\delta_{x,w_A(b,c)}M_{a|x}^A\otimes\delta_{y,w_B(c,a)}M_{b|y}^B\otimes \delta_{z,w_C(a,b)} M_{c|z}^C
\label{eq:lopf}
\end{align}
where $\delta_{i,j}$ is the Kronecker delta. Each output is sent to the process but also broadcasted in the global future as the outcome of the joint measurement. We refer to such measurement as ``LOPF'' (Local Operations with Process Function). When the process is causal-i.e., cannot violate causal inequalities-this reduces to standard LOCC, under the constraint of single-round protocols with fixed local projective measurements.

Consider that Alice, Bob and Charlie simply perform a projective measurement in the $\{\ket{0},\ket{1}\}$ (resp. $\{\ket{+},\ket{-}\}$) basis if they receive an input 0 (resp. 1), i.e. $\ket{a|x}=H^x\ket{a}$, $\ket{b|y}=H^y\ket{b}$ and $\ket{c|z}=H^z\ket{c}$, with $a,b,c,x,y,z\in\{0,1\}^6$, $H^{0}=\id$ denoting the identity and $H^{1}=H$ the Hadamard operator, such that their inputs are determined by the Lugano process, $w_{L}(a,b,c):\{w_A(b,c)=c(b\oplus 1)\hbox{ ; } w_B(a,c)=a(c\oplus 1) \hbox{ ; } w_C(a,b)=b(a\oplus 1)\}$.
This generates the SHIFT measurement, whose elements can be written as:
\begin{align}
    &E_{a,b,c}^{ABC}=H^{w_L(a,b,c)}\ketbra{a,b,c}{a,b,c}^{ABC}(H^{w_L(a,b,c)})^{\dagger} 
    \label{eq:shiftmeas}
\end{align}

\medskip

\subsection{SHIFT measurement with Quantum Switch} 
We now demonstrate that the SHIFT measurement can also be realized by local operations using \textit{coherently controlled classical communications}.

\begin{figure}[ht]
	\begin{center}
	\includegraphics[width=0.9\columnwidth]{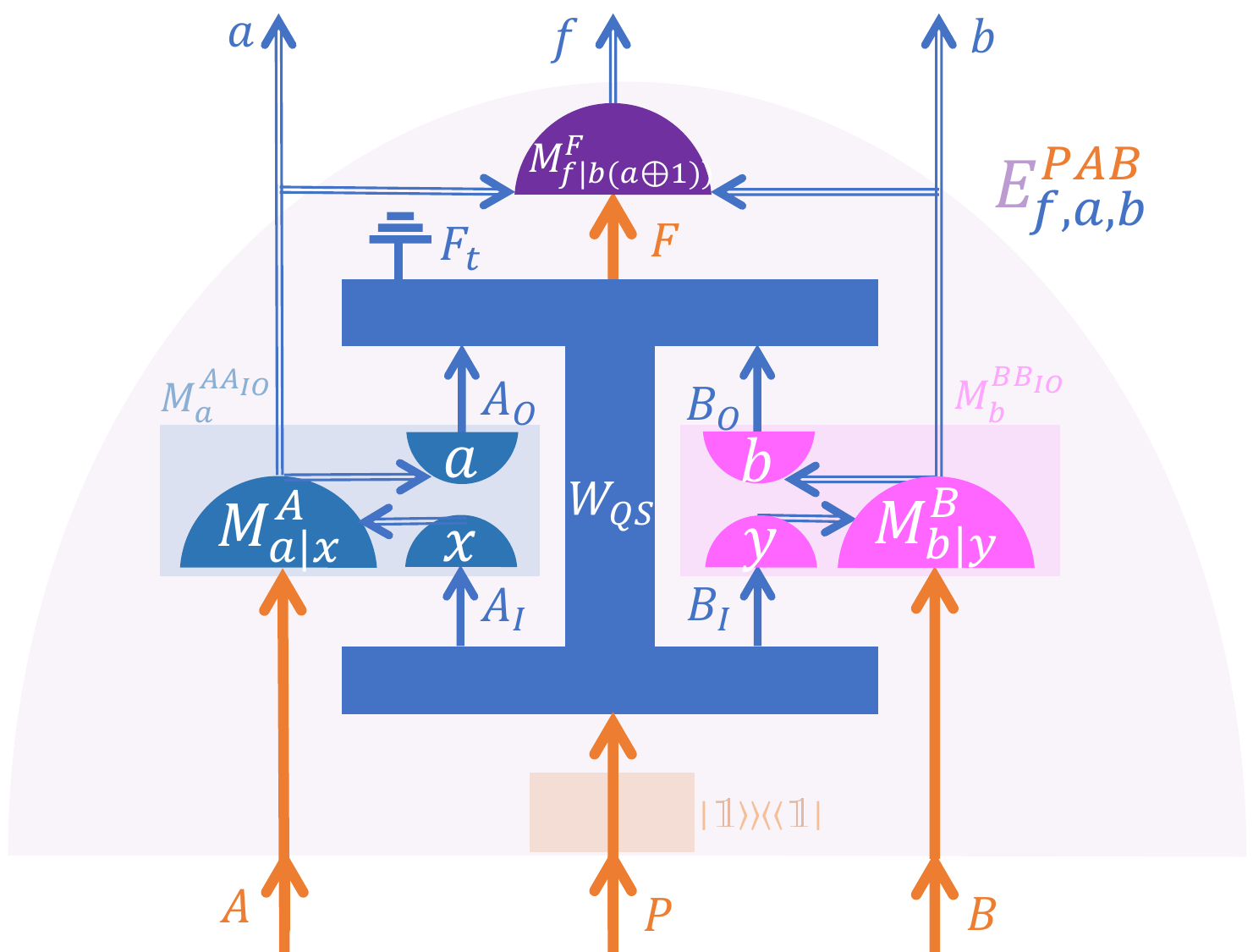}
	\end{center}
	\caption{The quantum switch realization of the SHIFT measurement: Alice and Bob perform classical measure and prepare operations on the process spaces that respectively define and are defined by their projective measurements $(M_{a|x}^A=\ketbra{a|x}{a|x}^A)_a$ and $(M_{b|y}^B=\ketbra{b|y}{b|y}^B)_b$ on their respective quantum systems (orange wires).  The order of their operations is controlled by the quantum system in $\HS^{PF}$ via a quantum switch (blue comb) $W_{QS}\in\mathcal{L}(\mathcal{H}^{PA_{IO}B_{IO}F})$ and Fiona's final measurement $(M_{f|a,b}^F=\ketbra{f|b(a\oplus 1)}{f|b(a\oplus 1)}^F)_f$, determined by the communication of $a$ and $b$ through side classical channels. This set-up generates an effective set of projectors $(E_{f,a,b}^{PAB})_{f,a,b}$, the SHIFT measurement.  } 
\label{fig:qs}
\end{figure}

Consider an alternative scenario where Alice and Bob control separate quantum labs with input and output Hilbert spaces $\HS^{A_I}$, $\HS^{A_O}$ for Alice, and $\HS^{B_I}$, $\HS^{B_O}$ for Bob. They also receive some auxiliary quantum states in Hilbert spaces $\HS^{A},\HS^{B}$. Each performs some fixed quantum operations described as quantum instruments~\cite{davies70}, i.e., sets of completely positive (CP) maps $\M_a: \L(\HS^{AA_I})\to\L(\HS^{A_O})$, $\M_b: \L(\HS^{BB_I})\to\L(\HS^{B_O})$, whose indices $a,b$ refer to some (classical) binary outcomes for Alice and Bob and whose sums $\sum_a \M_a$ and $\sum_b \M_b$   are trace-preserving (TP). Using the Choi isomorphism~\cite{choi75} (see Supplemental Material (SM) of \cite{dourdent21}), the CP maps $\M_a$, $\M_b$ can be represented as positive semidefinite (PSD) matrices $M_a^{AA_{IO}}$ and $M_b^{BB_{IO}}$.
Fiona, a third party in the causal future of Alice and Bob, controls a quantum lab with input Hilbert space $\HS^{F}$ but no output Hilbert space. She performs a measurement $(M_{f|a,b}^{F}\in\L(\HS^{F}))_f$ with (classical) outcome $f$. Fiona’s measurement setting $z$ is determined by the outcomes 
$a$ and $b$ from Alice and Bob, which are communicated to her via classical channels. A fourth party, Phil, in the causal past of all the other parties, acts trivially on a quantum input that he lets go through an identity channel in $\HS^{P}$. These operations are connected via a ``\textit{process matrix}'' $W^{PA_{IO}B_{IO}F} \in \L(\HS^{PA_{IO}B_{IO}F})$, a positive semidefinite matrix satisfying nontrivial linear constraints to generate valid probabilities~\cite{oreshkov1} (see e.g.~\cite{dourdent21}, SM, Sec.~B).
Within the process matrix framework \cite{oreshkov1}, this configuration defines an effective measurement $(E_{f,a,b}^{PAB})_{f,a,b}$  on the auxiliary quantum system in $\L(\HS^{PAB})$ (see \cite{dourdent24}, Appendix E) with: 
\begin{align}
&E_{f,a,b}^{PAB}=\notag\\ &\Tr_{A_{IO}B_{IO}F}\left[(M_{a}^{AA_{IO}}\otimes M_{b}^{BB_{IO}}\otimes M_{f|a,b}^{F})^T W^{PA_{IO}B_{IO}F}\right]
    \label{eq:dpovm}
\end{align}
Importantly, we also assume that all the involved Hilbert spaces are two-dimensional.
Furthermore, let us consider that the local operations performed by Alice, Bob, and Fiona are:
\begin{align}
     M_{a}^{AA_{IO}} & =\sum_x H^x\ketbra{a}{a}^{A}(H^x)^\dagger\otimes \ketbra{x}{x}^{A_{I}}\otimes   \ketbra{a}{a}^{A_O},\notag\\
      M_{b}^{BB_{IO}} & =\sum_y H^y\ketbra{b}{b}^{B}(H^y)^\dagger\otimes \ketbra{y}{y}^{B_{I}}\otimes  \ketbra{b}{b}^{B_O},\notag\\
       M_{f|a,b}^{F} &=H^{b(a\oplus 1)}\ketbra{f}{f}^{F}(H^{b(a\oplus 1)})^\dagger,
      \label{eq:qlocop}
\end{align}
Here, Alice’s operation projects in the computational basis on her input space $\HS^{A_I}$, retrieving the outcome $x\in\{0,1\}$. This outcome then sets up the projection $(H^x\ketbra{a}{a}^A(H^x)^\dagger)_a$ on her auxiliary system in $\HS^{A}$. The final outcome $a\in\{0,1\}$ is sent back into the process via $\HS^{A_O}$, and directly to Fiona via a classical channel. Bob’s operation follows a similar structure, while Fiona’s operation on $\HS^{F}$ depends on Alice and Bob’s outcomes $a$ and $b$ (Fig.~\ref{fig:qs}).  Importantly, Eq.\eqref{eq:qlocop} implies that we assume that Alice, Bob and Fiona communicate exclusively \textit{classical information} to each other.

The quantum switch (QS)~\cite{chiribella13} is the canonical example of a quantum process with indefinite causal order, which can be understood as a QC-QC~\cite{wechs1}. Its process matrix\footnote{The considered process is slightly different from the standard version of the quantum switch~\cite{chiribella13}. In the standard case, all communications between the parties are identity channels. Such standard quantum switch could also be used here, with a slight modification on Bob's operation (Eq.~\eqref{eq:qlocop}), who would then need to send $\ketbra{b\oplus 1}{b\oplus 1}^{B_O}$ back in the process instead of $b$.} can be written as:
\begin{align}
   & W_{QS}=\Tr_{F_t}\ket{w_{QS}}\bra{w_{QS}},\notag\\
&\ket{w_{QS}}=\ket{0}^{P}\otimes\ket{0}^{F}\otimes\ket{0}^{A_{I}}\otimes\dket{\id}^{A_{O}B_I}\otimes\dket{\id}^{B_OF_t}\notag\\
&+\ket{1}^{P}\otimes\ket{1}^{F}\otimes\ket{0}^{B_{I}}\otimes\dket{X}^{B_{O}A_I}\otimes\dket{X}^{A_OF_t}
\label{eq:qs}
\end{align}
where $\dket{\id}^{XX'} =  \ket{0}^{X}\otimes\ket{0}^{X'}+\ket{1}^{X}\otimes\ket{1}^{X'}$ ;\\
$\dket{X}^{XX'} = \ket{0}^{X}\otimes\ket{1}^{X'}+\ket{1}^{X}\otimes\ket{0}^{X'}$.\\

Here, the causal order between Alice and Bob's operations on a ``target'' system (initially in state $\ket{0}$) is coherently controlled by a qubit in $\mathcal{H}^{PF}$. If this ``control'' is $\ket{0}$, the target goes to Alice and then Bob (it is then routed to the final target space $\mathcal{H}^{F_t}$, which is traced out). If the control is $\ket{1}$, the order is reversed. For a superposition control state $\ket{\pm}$, these two causal orders interfere. $W_{QS}$ is ``\textit{causally non-separable}'' \cite{oreshkov16,araujo1}: it cannot be decomposed as a convex mixture of fixed-order processes.  \\

The elements of the effective measurement (Eq.~\eqref{eq:dpovm}) generated by the implementation of Alice, Bob and Fiona's local operations (Eq.~\eqref{eq:qlocop}) within the quantum switch ($W^{PA_{IO}B_{IO}F}=W_{QS}$ from Eq.~\eqref{eq:qs}) are:
\begin{align}
    E_{f,a,b}^{PAB} =\delta_{f,0}\delta_{b(a\oplus 1),0}\ketbra{f}{f}^{P}\ketbra{a}{a}^A(H^a\ketbra{b}{b}(H^a)^\dagger)^B\notag\\+\delta_{f,1}\delta_{b(a\oplus 1),0}\ketbra{f}{f}^{P}(H^{b\oplus 1}\ketbra{a}{a}(H^{b\oplus 1})^\dagger)^A\ketbra{b}{b}^B\notag\\
    +\delta_{b(a\oplus 1),1}(H\ketbra{f}{f}H^\dagger)^{P}\ketbra{a}{a}^A\ketbra{b}{b}^B.
    \label{eq:shiftqs}
\end{align}
It is straightforward to verify that these are the elements (Eq.~\eqref{eq:shiftmeas}) of the SHIFT measurement.

The first term of Eq.~\eqref{eq:shiftqs} indicates that if $f=0$ and $b(a\oplus 1)=0$, Alice measures in the computational basis, while Bob's choice of basis depends on her outcome $a$. Thus the SHIFT projectors on $\{\ket{000},\ket{01+},\ket{01-}\}^{PAB}$ are elements of Eq.~\eqref{eq:shiftqs}. The second term indicates that if $f=1$ and $b(a\oplus 1)=0$, Bob measures in the computational basis, while Alice's basis choice depends on  $b\oplus 1$, yielding the SHIFT projectors on $\{\ket{111},\ket{1+0},\ket{1-0}\}^{PAB}$. The system in $\mathcal{H}^P$ controls their causal order. Finally, whenever $b(a\oplus 1)=1$, Alice and Bob always output $a=0$ and $b=1$, producing the SHIFT projectors on $\{\ket{+01},\ket{-01}\}^{PAB}$, which encode no causal order -- making it ``indefinite''\footnote{$\ket{01}$ is common to the two bipartite bases: ${\ket{00},\ket{01},\ket{1+},\ket{1-}}$, where Alice is in the causal past of Bob, and ${\ket{+0},\ket{01},\ket{-0},\ket{11}}$, where Bob is in the causal past of Alice.}.
\medskip

\subsection{From the Lugano process to the quantum switch}\label{sec:LuganoToQS}
  The Lugano process is a classical noncausal resource capable of maximally violating the causal inequality  $P((x,y,z)=w_{L}(a,b,c))\leq 3/4$. This violation is a
device-independent (DI) certification, where the causal non-separability is certified by the observed statistics $P(x,y,z|a,b,c)$, in a scenario where Alice, Bob and Charlie only receive and produce classical bits. In contrast, the quantum switch, despite its causal non-separability, is inherently a causal process. As a QC-QC \cite{wechs1}, it cannot violate such causal inequalities. This occurs because Fiona, being in the causal future, does not signal to any party, and that the marginal probability distribution of Alice and Bob remains causal \cite{araujo1,oreshkov16}.

 On the other hand, the certification of the causal non-separability of the quantum switch  can occur in a semi-DI scenario, where the parties perform uncharacterised operations on trusted quantum inputs  (SDI-QI) \cite{dourdent21,dourdent24}. The realization of the SHIFT measurement can be expressed as maximally winning a state identification game in such a SDI-QI scenario, where Alice, Bob and Charlie must recover the label $(\alpha,\beta,\gamma)$ of each SHIFT state, with the (SDI-QI) causal inequality:
\begin{equation}
\resizebox{\columnwidth}{!}{$\displaystyle
\frac{1}{8} \sum_{\alpha,\beta,\gamma} P(a=\alpha,b=\beta,c=\gamma |\ket{\psi}_{\alpha,\beta,\gamma}^{ABC}=H^{w_{L}(\alpha,\beta,\gamma)}\ket{\alpha,\beta,\gamma} )\leq \xi$}
\end{equation}
where $\xi<1$ is the causal bound, and 1 the algebraic maximum attainable with a SHIFT measurement\footnote{Assuming the measurement is implemented in a causal LOPF scenario, $\xi=7/8$. Indeed there is always a well-defined global past in a causal process function. This means that one of the inputs, $(x,y,z)$, must be constant, e.g. set to 0. Then, at most six out of the eight SHIFT states can be perfectly distinguished, while the remaining two must be guessed randomly.}  (see Appendix \ref{app:cnsshift}, Eq.~\eqref{eq:p2fineq}).
Here, causal nonseparability is thus certified by the 
distributed measurement $(E_{a,b,c}^{ABC})_{a,b,c}$ (or $(E_{f,a,b}^{PAB})_{f,a,b}$ after relabeling), acting on quantum inputs$(\ket{\psi}_{\alpha,\beta,\gamma}^{ABC})_{\alpha,\beta,\gamma}$ (or $(\ket{\psi}_{\gamma,\alpha,\beta}^{PAB})_{\gamma,\alpha,\beta}$). Implementing locally the SHIFT measurement offers a physically intuitive algebraic maximal witness of (SDI-QI) causal nonseparability for both the Lugano process and the QS. 

Eventually, a difference between the two SHIFT measurement constructions emerges in some distributed guessing problems, like quantum data hiding \cite{divincenzo02,kunjwal23a}:  in the Lugano process, the local outputs remain hidden and each party guesses others' outputs correctly one time in four (for input 1); while in the QS (Eq.~\eqref{eq:qs}), Fiona receives $a$ and $b$ separately. Nevertheless, in Appendix \ref{app:lugqcqc}, we show how a Lugano process matrix \cite{araujo4} can become a causally non-separable QC-QC via a simple SWAP operation, embedding the classical side channels from Alice and Bob to Fiona (Fig.~\ref{fig:qs}) into the process. Simplifying the discrimination game to one quantum input (the control one), we derive in Appendix \ref{app:ndi}, the following inequality, analogous to the standard Lugano one:
\begin{equation}
    \resizebox{\columnwidth}{!}{$\displaystyle
    \frac{1}{8} \sum_{a,b,\gamma} P(f=\gamma,(x,y,z)=w_L(a,b,\gamma)|\ket{\psi}^P_{\gamma|a,b}=H^{b(a\oplus 1)}\ket{\gamma})\leq \xi$}
\end{equation}

This derivation can also be understood through the realization of the Lugano process on time-delocalized subsystems~\cite{Wechs23}, namely as a quantum circuit in which certain quantum systems are delocalized in time. In this representation, Alice’s and Bob’s variables are time-delocalized in a straightforward manner, being classically controlled by Charlie’s process systems. Charlie’s systems, however, are time-delocalized in a more subtle way: Charlie’s operation is applied both at the beginning of the circuit and may later be reversed and reapplied at the end, conditioned on the outcomes of Alice’s and Bob’s operations (Fig.~\ref{fig:timedeloc}). Importantly, this should not be interpreted as Charlie’s operation being performed multiple times; rather, it acts only once on time-delocalized input and output subsystems.

This classically controlled switch between Alice and Bob, mediated by a time-delocalized Charlie, can be converted to a genuine quantum switch by endowing Charlie with an auxiliary quantum system that serves as the quantum control of Alice’s and Bob’s causal order. Moreover, by a fine-graining of agentivity—splitting Charlie into two distinct parties, Phil and Fiona, located respectively in the global past and future of Alice and Bob—the resulting causal structure naturally takes the form of a QC-QC (Fig.~\ref{fig:causalgraph}).

The concept of time-delocalized subsystems was originally introduced in the context of an ongoing debate concerning the interpretation of experimental realizations of the quantum switch, predominantly in photonic platforms \cite{procopio,rubino,goswami18,wei19,rubino22,goswami20,guo20,taddei20,rubino21,cao23,stromberg23,antesberger24} (see \cite{rozema24} for a review). At the heart of this debate lies the question of whether such “photonic quantum switch” experiments genuinely instantiate indefinite causal order, or whether they should instead be regarded as simulations of phenomena that would only arise from superpositions of gravitational causal structures, i.e., a “gravitational quantum switch” \cite{maclean17,oreshkov19,paunkovic20,ormrod23,vilasini24,vilasini24a}.

The stance one adopts in this debate, mostly based on a chosen notion of event, directly affects the interpretation of any possible experimental local realization of the SHIFT measurement via a quantum switch, and hence of a simulation of the Lugano process. In Appendix~\ref{app:exp}, after briefly reviewing this controversy, we examine in greater detail how our transformation of the Lugano process into a QC-QC fits within the framework of time-delocalized subsystems. We also propose a concrete photonic implementation of the SHIFT measurement based on a quantum switch realized as a “causal loop with an undetermined cut” within a Mach–Zehnder interferometer (Fig.~\ref{fig:shiftexp}). Finally, we discuss the potential implications of a gravitational realization of the SHIFT measurement for the notion of causal indefiniteness.

\medskip

\subsection{LOPF vs. LOSupCC}\label{sec:LOPF vs. LOSupCC}

The main difference between the SHIFT measurement from the Lugano process \cite{kunjwal23a} and the QC-QCs is that the latter -- despite involving solely exchange of classical information --  no longer occur within a genuinely tripartite LOPF scenario. As shown in Fig.~\ref{fig:qs} and Fig.~\ref{fig:qcqc} in Appendix, unlike Alice and Bob, Fiona does not have direct access to a SHIFT substate. Instead, she receives the system prepared in $\HS^P$ only after it passes through the effective map (in Choi representation): $C_{a,b}^{PF}=(\rho^A\otimes \rho^B\otimes W_{QS})*(M_{a}^{AA_{IO}}\otimes M_{b}^{BB_{IO}})$, which depends on Alice and Bob's local operations. One might argue that Alice and Bob indirectly communicate to Fiona via this quantum channel. However, if $\rho^A\otimes\rho^B$ are projectors on $\{\ket{00},\ket{1\pm},\ket{\pm 0},\ket{11},\ket{01}\}$ 
with the operations Eq.\eqref{eq:qlocop}, this channel is always operationally indistinguishable from an identity channel, as it always perfectly routes the remaining adequate SHIFT substate $\rho^P$ to Fiona. Aside from this quantum control transformation, all other involved communications, encoded in quantum systems within $\HS^{A_{IO}B_{IO}F_t}$, are classical. Furthermore, the QC-QC-SHIFT can be represented without requiring quantum encoding of the classical communications:
\begin{equation}
    \resizebox{\columnwidth}{!}{$\displaystyle
    E_{f,a,b}^{PAB}=\delta_{z,b(a\oplus 1)}H^{z}\ketbra{f}{f}^{P}(H^z)^\dagger\otimes M_{a|w_A^f(b)}^A\otimes M_{b|w_B^f(a)}^B$}
\end{equation}
where $w_A^{f}(b)=f(b\oplus1)$ and $w_B^{f}(a)=a(f\oplus1)$ are the reduced Lugano functions generating $x$ and $y$ respectively from a fixed $f$. Consequently, the SHIFT measurement can be realized via two \textit{local operations with quantum control (or superposition) of classical communications (LOSupCC)}, in a $(P+2+F)$-partite scenario. Importantly, this  does not refute the established correspondence between the SHIFT measurement and the Lugano process, but clarifies that successful SHIFT discrimination is ``an operational signature of noncausality'' \cite{kunjwal23a}  \textit{only when assuming an LOPF scenario}. In a different scenario such as the $(P+2+F)$-partite case, it only witnesses the weaker notion of causally non-separable measurement. \\
The perfect routing of the control system to Fiona is a key feature here, 
and is also leveraged in the ``DRF-DI certifications'' of the QS \cite{lugt23,lugt24, gogioso23}. This suggests a link between the DRF inequality violated by the quantum switch \cite{lugt23} with the causal inequality violated by the Lugano process. However, unlike the DRF  and the standard DI-certification scenarii, this approach is theory-dependent, relying on quantum theory and the process matrix formalism. 
Nevertheless, this reveals a new class of ``SupCC'' processes (beyond the standard QS) that exhibit SDI-QI causal nonseparability, and can be certified in a Network-DI scenario \cite{dourdent24} (see Appendix \ref{app:ndi}), via intuitive NLWE-bases witnesses.
\medskip

\textit{Theorem}--- A $N-$partite NLWE basis $\{H^{w(\bm{a})}\ket{\bm{a}}\}_{\bm{a}=(a_1,...,a_N)}$ measurable by local operations with a Boolean process function without global past $w$ satisfying the
transparent control channel condition 
\begin{align}
    \exists i, \hbox{ s.t. } \forall \bm{a}_{\backslash i} \hbox{ s.t. } w_i(\bm{a}_{\backslash i})=1, \notag\\
    w^{a_i=0}(\bm{a}_{\backslash i})=w^{a_i=1}(\bm{a}_{\backslash i})
    \label{eq:transparent}
\end{align}
is also measurable by  LOSupCC.\medskip

The proof is provided in Appendix \ref{app:theorem}.
Whether all Boolean process functions without a global past satisfy the transparency condition remains an open question. Based on our preliminary analysis of various four-partite cases (Appendix \ref{app:4partite}), each exhibiting distinct noncausal structures and satisfying this condition \cite{araujo3,baumeler22,tobar}, we conjecture that the answer is affirmative.
This result offers a new perspective on the noncausality of these process functions, which can be seen as arising from the combination of a ``causally non-separable channel'' - implementable from a physical causally non-separable process (QC-QC) -  and a simple loop that feeds the classical outputs of local operations back as inputs to the channel. It also opens the door to defining and characterizing a new hierarchy of (deterministic) classical communications in the spirit of \cite{wechs1}, and exploring novel indefinite-causal-order-based advantages for information-processing tasks beyond the standard LOCC paradigm.\\

\begin{acknowledgments}
We thank Antonio Ac\'in, Ognyan Oreshkov, Raphaël Mothe, Alastair Abbott, Cyril Branciard, Ravi Kunjwal, Elie Wolfe, Gustavo Balvedi Pimentel, Victoria Wright, Marco Túlio Quintino, Časlav Brukner for enlightening discussions, Alexei Grinbaum, Giorgio Trespidi and an anonymous reviewer of QPL 2025 for feedbacks on a first manuscript, and acknowledge financial support from the EU NextGen Funds, the Government of Spain (FIS2020-TRANQI and Severo Ochoa CEX2019-000910-S, PRE2022-101475), Fundació Cellex, Fundació MirPuig, Generalitat de Catalunya (CERCA program).

\end{acknowledgments}

\bibliography{bibli}

\appendix

\section{Simulating the Lugano process with the SHIFT measurement}\label{app:lugshift}
To demonstrate how the Lugano process can be derived from the SHIFT measurement, we begin with an intuitive idea. Each element of the SHIFT basis can be associated with an event of the form $(x,y,z\vert a,b,c)$ where $a,b,c,x,y,z\in\{0,1\}^6$. Here, $a,b$ and $c$ represent the values of the three bits encoded in the SHIFT state, while $x,y$ and $z$ denote the basis in which each bit is measured. For instance, the states $\ket{000}$ and $\ket{01+}$, correspond to the events $(000\vert000)$ and $(001\vert010)$, respectively. It can be observed that all events constructed this way satisfy the conditions $x=c(b\oplus1)$, $y=a(c\oplus1)$ and $z=b(a\oplus1)$, hence the Lugano process.

\begin{figure}[ht]
	\begin{center}
	\includegraphics[width=0.95\columnwidth]{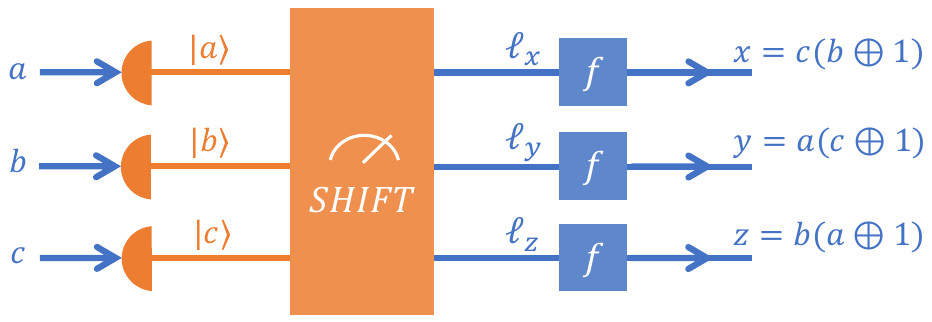}
	\end{center}
	\caption{Circuit implementing the classical channel underlying the Lugano process from the SHIFT measurement. (Fig. 2 from \cite{kunjwal23a}). This circuit can be interpreted as a simulation of the noncausal Lugano process, which would be genuinely implemented if $x,y,z$ were respectively in the local pasts of $a,b,c$.  } 
	\label{fig:shiftlug}
\end{figure}

We now present the more rigorous approach from \cite{kunjwal23a} to deriving these events, demonstrating how the classical channel underlying the  Lugano process can be implemented using a device that measures an incoming state in the SHIFT basis (Fig.~\ref{fig:shiftlug}). Alice, Bob, and Charlie encode a classical bit $a,b,c \in \{0,1\}^3$ in the computational basis, preparing the state $\ket{abc}$, which is then sent to a SHIFT measurement device. The device outputs a classical variable for each input, denoted as $\ell_x,\ell_y,\ell_z\in\{0,1,+,-\}^3$.  Finally, to obtain the Lugano process, a function $f$ is applied to the measurement outcomes, defined as $f(0)=0$, $f(1)=0$, $f(+)=1$ and $f(-)=1$. As a result we obtained a process which takes three input bits $a,b,c$ and, by measuring the elements of the SHIFT basis, produce three bits $x=f(\ell_x)=c(b\oplus1)$, $y=f(\ell_y)=a(c\oplus1)$ and $z=f(\ell_z)=b(a\oplus1)$.
As an example, consider the case where Alice, Bob, and Charlie provide identical input bits $a=b=c\in\{0,1\}$. Since the prepared state belongs to the SHIFT basis, the measurement outcomes are necessarily $\ell_x=\ell_y=\ell_z\in\{0,1\}$.  Applying the function $f$ then yields $x=y=z=0$, as expected from the Lugano process. 
Now consider a scenario where one input bit differs from the others, for instance $a=c=0$ and $b=1$. In this case, the encoded state is $\ket{010}$, which does not belong to the SHIFT basis. Consequently, the measurement device can output
$(\ell_x,\ell_y,\ell_z)=(0,1,+)$ or $(0,1,-)$ each with probability $\vert \braket{01+}{010}\vert^2=\vert \braket{01-}{010}\vert^2=1/2$. In both cases, applying the function $f$, results in $x=y=0$ and $z=f(+)=f(-)=1$, thus correctly reproducing the expected Lugano process output.

\medskip

\section{Converting the Lugano process into a QC-QC}
\label{app:lugqcqc}
From (Theorem 4, \cite{baumeler22}), all (here tripartite) process functions can be written as a process matrix:
\begin{align}
W=\sum_{a,b,c}\ket{a,b,c}\bra{a,b,c}^{A_OB_OC_O}\otimes \ket{x,y,z}\bra{x,y,z}^{A_IB_IC_I}
\end{align}
where $x:=w_A(b,c)$, $y:=w_B(a,c)$ and $z:=w_C(a,b)$.  
\\

It was proven in (Theorem 2, \cite{baumeler19}) that every process function can be purified, i.e. can be extended to a reversible one. From \cite{araujo4}, the purification procedure consists in using the standard method to make an irreversible function reversible, which involves transforming the function from \( \ket{x} \to \ket{f(x)} \) to \( \ket{x}\ket{y} \to \ket{x}\ket{y \oplus f(x)} \). For our specific case, this corresponds to mapping \( \ket{a, b, c}^{A_OB_OC_O}\ket{i, j, k}^{P_AP_BP_C} \) to \( \ket{a, b, c}^{F_AF_BF_C}\ket{i \oplus x, j \oplus y, k \oplus z}^{A_IB_IC_I} \). The purification of $W$ is thus given by $W_{rev} =\ketbra{w}{w}_{rev}$ with
\begin{align}
  &\ket{w}_{rev} = \sum_{a,b,c,i,j,k} \ket{i,j,k}^{P_AP_BP_C}\ket{a,b,c}^{F_AF_BF_C}\notag\\
   &\otimes\ket{i\oplus x,a}^{A_{IO}}\ket{j\oplus y,b}^{B_{IO}}\ket{k\oplus z,c}^{C_{IO}}
\end{align}
The relation between $W_{rev}$ and $W$ is given by
\begin{align}
    W=\ketbra{000}{000}^{P_AP_BP_C}*W_{rev}*\id^{F_AF_BF_C}
\end{align}
$*$ is the ``\textit{link product}''~\cite{chiribella08,chiribella09}, a convenient tool for calculations defined for any matrices $M^{XY} \in \L(\HS^{XY})$, $N^{YZ} \in \L(\HS^{YZ})$ as $M^{XY}*N^{YZ} = \Tr_Y[(M^{XY}\otimes\id^Z)^{T_Y}(\id^X\otimes N^{YZ})] \in \L(\HS^{XZ})$ (where $T_Y$ is the partial transpose over $\HS^Y$; see also~\cite{dourdent21}, SM, Sec.~A). (A full trace $\Tr[(M^Y)^T N^Y]$ and a tensor product $M^X \otimes N^Z$ can both be written as a link product. Moreover, the link product is commutative and associative.)

Let us now consider a partial purification of $W$, in which only the global future of one party, e.g. Charlie\footnote{In general, this choice cannot be arbitrary, as shown in (Section \ref{app:4partite}). However, in the specific case of the Lugano process, because it is perfectly symmetric, the choice does not matter here.}, is left open.
\begin{align}
    W_{prev}=\ketbra{000}{000}^{P_AP_BP_C}*W_{rev}*\id^{F_AF_B}
    \label{eq:pp}
\end{align}
$W_{prev}\in\L(\HS^{A_{IO}B_{IO}C_{IO}F_C})$. We can identify the
global future associated with Charlie as the global future of the process, $\HS^{F_C}\equiv\HS^{F}$. 

Finally, consider Charlie's quantum instrument $(M_c^{CC_{IO}})_c$. In order to realize the SHIFT measurement, Charlie performed the operations
\begin{equation}
    M_c^{CC_{IO}}=\sum_z H^z\ketbra{c}{c}^{C}(H^z)^\dagger\otimes \ketbra{z}{z}^{C_{I}}\otimes   \ketbra{c}{c}^{C_O}
\end{equation}
i.e. Charlie received his setting $z$ from the process, performed the adequate measurement on his auxiliary system $\HS^C$, and sent the outcome $c$ back to the process. Let us now assume that instead of the previous strategy, Charlie still performs a computational basis measurement on his process input to retrieve $z$, but he sends his auxiliary quantum input directly into the process via an identity channel instead of measuring it:
\begin{align}
    M_z^{CC_{IO}}= \dketbra{\id}{\id}^{CC_O}\otimes \ketbra{z}{z}^{C_{I}}
\end{align}
These two sub-operations—the identity channel and the computational basis measurement—are independent of one another. Instead of assuming they are both performed within Charlie's single laboratory, let us consider an alternative scenario where they are implemented separately in distinct laboratories. 
 
Specifically, we introduce a global past laboratory, Phil's, where the identity channel is applied. Here, we assume that \(\mathcal{H}^C\) and \(\mathcal{H}^{C_O}\) are isomorphic, allowing us to use an abusive notation and relabel them as \(\mathcal{H}^C \equiv \mathcal{H}^{C_O} \equiv \mathcal{H}^P\). Additionally, we introduce a global future laboratory, Fiona's, where the measurement $(M_f^{FF_t})_f$ is performed:
\begin{align}
M^P&=\dketbra{\id}{\id}^P\notag\\
    M_{f}^{FF_t}  &=\sum_z(H^z\ketbra{f}{f}(H^z)^\dagger)^{F}\otimes\ketbra{z}{z}^{F_t}
    \label{eq:qclop}
\end{align}
where we relabeled $\HS^{C_I}\equiv\HS^{F_t}$, and  $f$ replaces $c$.

\begin{figure}[ht]
	\begin{center}
	\includegraphics[width=0.5\textwidth]{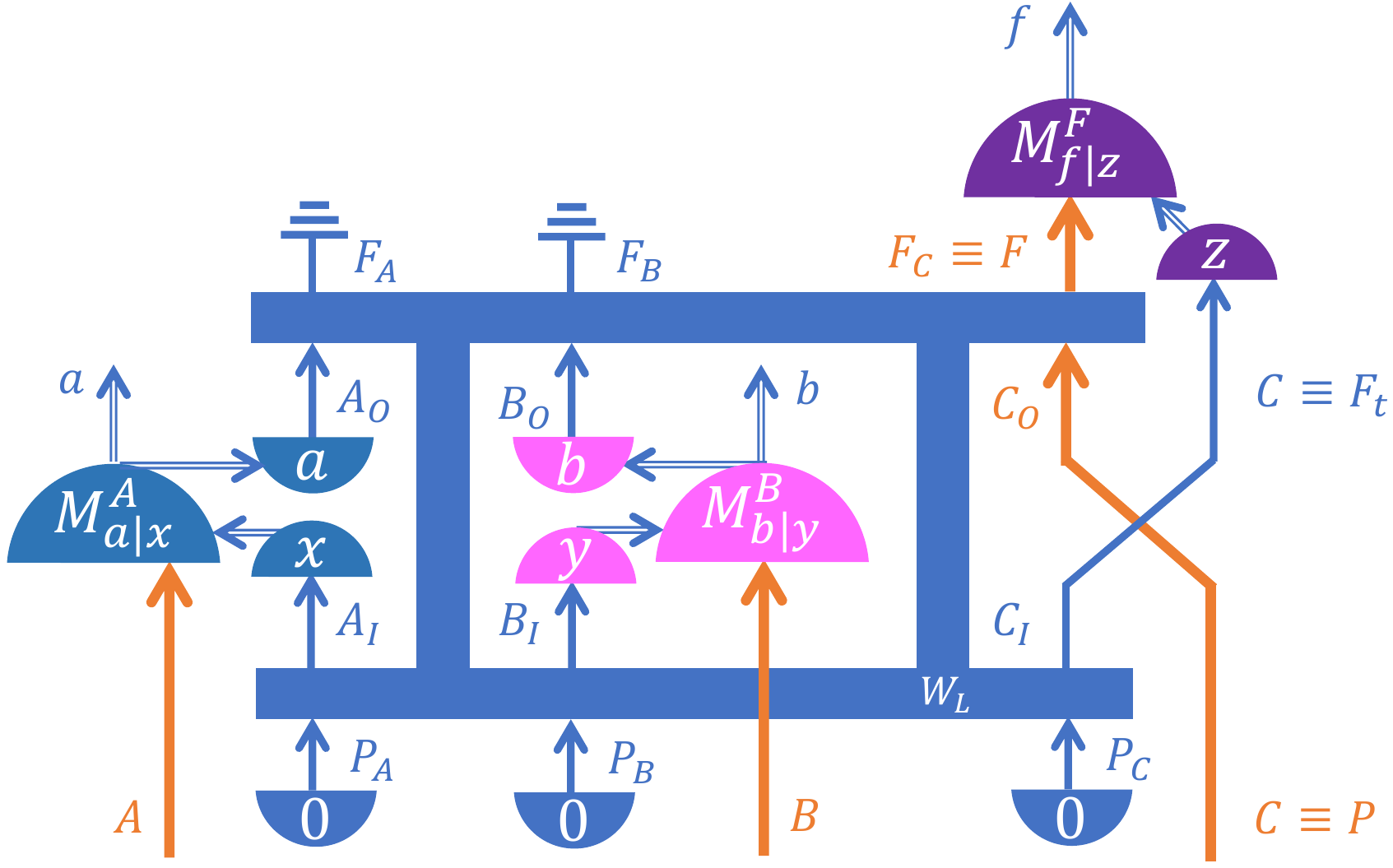}
	\end{center}
	\caption{From the Lugano process to the QC-QC: All the global past spaces of the purified Lugano process $W_L$ are initialized in state $\ket{0}^{P_X}$, and the systems in $\HS^{F_A}$ and $\HS^{F_B}$ are traced out. Alice and Bob operations are unchanged: $M_{a|x}^A=\ketbra{a|x}{a|x}^A$ with $\ket{a|x}=H^x\ket{a}$ and $M_{b|y}^B=\ketbra{b|y}{b|y}^B$ with $\ket{b|y}=H^y\ket{b}$. The global future from the purification of Charlie's systems is left open, i.e. is identified as the global future of the process, $\HS^{F_C}\equiv\HS^{F}$. Charlie performs a $SWAP$ between his process systems and his auxiliary systems. His auxiliary input system, i.e. the SHIFT substate he receives, can be identified as the global past of the process, $\HS^{C_O}\equiv \HS^{P}$. The input received from the process is sent to Fiona $\HS^{C_I}\equiv \HS^{F_t}$, who measures it in the computational basis, giving outcome $z$. Fiona then measures the process global future system following $M_{f|z}^F=\ketbra{f|z}{f|z}^F$ with $\ket{f|z}=H^z\ket{f}$.  } 
	\label{fig:lugqs}
\end{figure}

Equivalently, we can also consider $\HS^{C}=\HS^{C_{(I)}}\otimes\HS^{C_{(O)}}$ with $dim(\HS^{C_{(O)}})=dim(\HS^{C_{(I)}})=2$, such that $\HS^{C_{(I)}}\equiv\HS^{P}$ and $\HS^{C_{(O)}}\equiv\HS^{F_t}$, and such that Charlie performs a  SWAP operation between its four systems: the global past wire in $\HS^{P}$ is pulled in the process input space $\HS^{C_O}$, while the process output wire $\HS^{C_I}$ is pulled to the global future $\HS^{F_t}$ (Fig.~\ref{fig:lugqs}). When this SWAP operation is implemented in the partially purified Lugano process  Eq.~\eqref{eq:pp} (or alternatively, considering Charlie's splitting into Phil and Fiona) with $x,y,z$ defined by Eq.~\eqref{eq:lugano}, we obtain the process matrix
\begin{align}
    W_{QCQC}&=W_{prev}^{Lugano}*(\dketbra{\id}{\id}^{PC_O}\otimes\dketbra{\id}{\id}^{C_IF_t})\notag\\&=\ket{w_{QCQC}}\bra{w_{QCQC}}, \hbox{  with  }\notag\\
\ket{w_{QCQC}}&=\notag\\
&\hspace{-8mm}\sum_{a,b}(\ket{00}^{PF}\otimes\ket{0}^{A_{I}}\otimes\ket{a,a}^{A_OB_{I}}\otimes\ket{b,b(a\oplus 1)}^{B_O F_t}\notag\\
&\hspace{-13mm}+\ket{11}^{PF}\otimes\ket{0}^{B_{I}}\otimes\ket{b,(b\oplus 1)}^{B_{O}A_I}\otimes\ket{a,b(a\oplus 1)}^{A_O F_t})
\label{eq:qcqc}
\end{align}

Eq.~\eqref{eq:qcqc} is a quantum circuit with quantum control of causal orders (QC-QC) \cite{wechs1} similar to a quantum switch Eq.~\eqref{eq:qs} in which Fiona receives a function of both Alice and Bob's outcomes as an input ($z=b(a\oplus 1)$).

\begin{figure}[ht]
	\begin{center}
	\includegraphics[width=0.95\columnwidth]{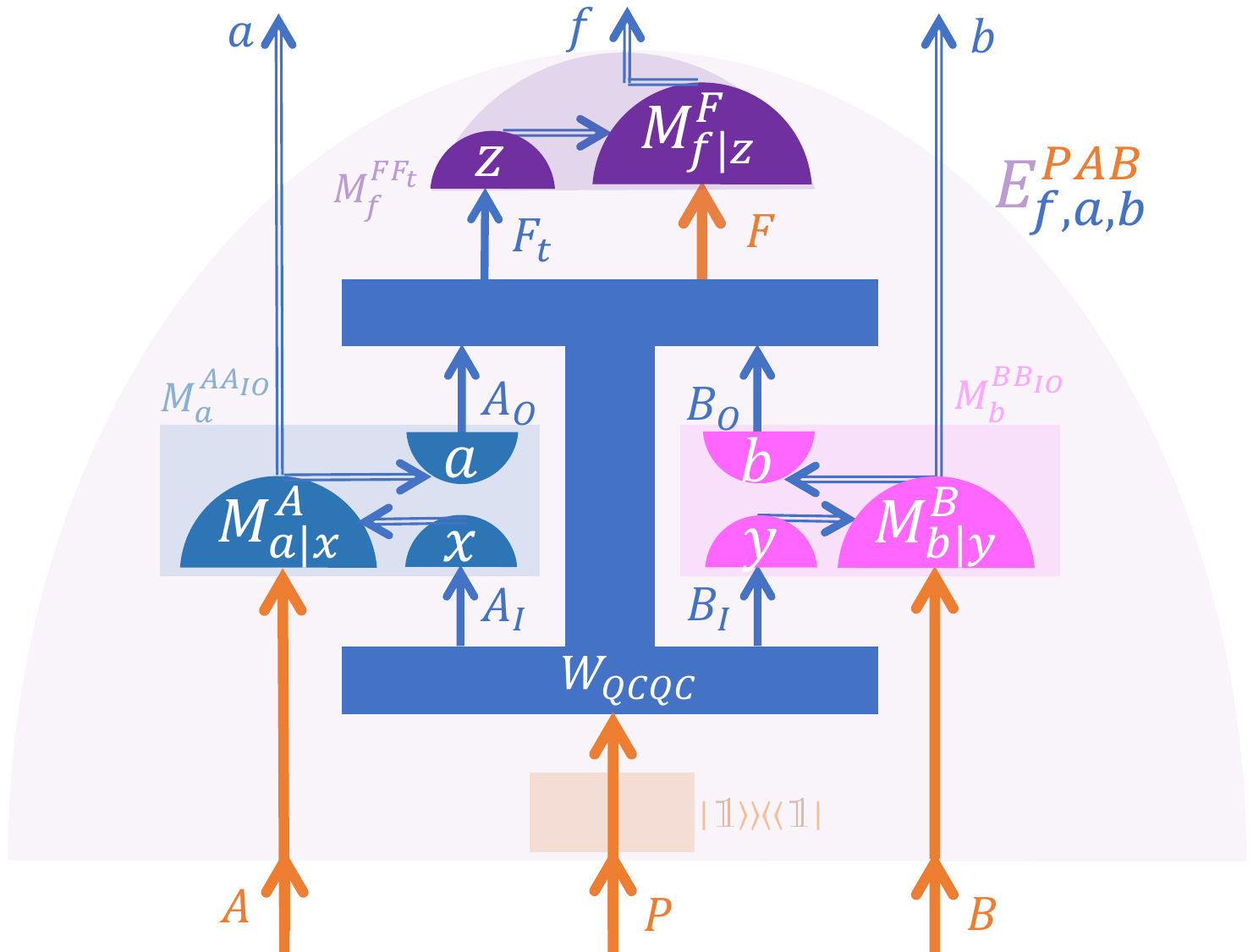}
	\end{center}
	\caption{QC-QC realization of the SHIFT measurement } 
	\label{fig:qcqc}
\end{figure}

While the conversion from a process function to a QC-QC is a general technique not unique to the Lugano process, its application here yields an important insight. The conversion preserves the indefinite causal structure that enables the SHIFT measurement (Fig.~\ref{fig:qcqc}). This preservation reveals a new connection between the Lugano process and the QS, illuminating why both resources can successfully perform this measurement despite their fundamental difference.

\section{Proof of Theorem 1}\label{app:theorem}
Consider a $N-$partite Boolean process function $w:\{0,1\}^{N}\rightarrow \{0,1\}^N $, a set of $N$ Boolean functions characterized by a unique fixed point condition \cite{baumeler16,tobar} in order to preserve logical consistency under free local operations. One of the consequences of this condition is that of non-self-signalling, i.e. for each party, their input is
independent of their own output: $w_i(\bm{a})=w_i(\bm{a}_{\backslash i})$, where $\bm a$ is the set of outputs of all parties and $\bm{a}_{\backslash i}$
is the set of outputs of all parties except the
ith party. Thus, $w(\bm{a}):\{w_i(\bm{a}_{\backslash i})\}_{i=1}^N$. Let us consider the special case of process functions \textit{without global past},  where each party can receive a signal from at least one other party,  
\begin{align}
    \forall i, \exists k, \bm{a}_{\backslash i}\in\{0,1\}^{N-1}: w_i(\bm{a}_{\backslash i})\neq w_i(\bm{a}_{\backslash i}^{(k)})
\end{align}
with $\bm{a}_{\backslash i}^{(k)}=(a_1,...,a_{k-1},a_k\oplus 1,a_{k+1},...,a_n)$. The Lugano process Eq.~\eqref{eq:lugano} is an example of such a Boolean process function without global past.\\

It was shown in \cite{kunjwal23a}
that if $w$ is a $N-$partite Boolean process function without global past, then $\{H^{w(\bm{a})}\ket{\bm{a}}\}_{\bm{a}}$ is a NLWE basis, which can thus be measured by local operations and the corresponding process function. Let us show that if $w$ satisfies the transparent control channel condition \begin{align}
    \exists i, \hbox{ s.t. } \forall \bm{a}_{\backslash i} \hbox{ s.t. } w_i(\bm{a}_{\backslash i})=1, \notag\\
    w^{a_i=0}(\bm{a}_{\backslash i})=w^{a_i=1}(\bm{a}_{\backslash i})
    \label{eq:transparentb}
\end{align}
then the associated NLWE measurement can also be measured by local operations with superposition of classical communications.\\

From (Theorem 4, \cite{baumeler22}), all (here Boolean) process functions can be written as a process matrix:
\begin{align}\nonumber
W=&\sum_{\bm a}\bigotimes_{j=1}^N\ketbra{ a_j}{ a_j}^{{A_{O}^{(j)}}}\otimes \ketbra{x_j}{x_j}^{{A_{I}^{(j)}}}.
\end{align}

with $x_j:=w_j(\bm a_{\backslash j})$, $W\in\L(\bigotimes_{j=1}^N \HS^{A_{IO}^{(j)}})$, where $\HS^{A_{I}^{(j)}}$ and $\HS^{A_{O}^{(j)}}$ denote respectively the input and output Hilbert spaces of party $j$.

Following the procedure used in Section~\ref{app:lugqcqc},  
we consider a partial purification of $W$, leaving open only the global future of the party, denoted by the label $i$, that satisfies the transparency condition~\ref{eq:transparentb}. The resulting purified process is $W_{prev}=\ketbra{w_{prev}}{w_{prev}}$, where:
\begin{align}
    \ket{w_{prev}}=\sum_{\bm{a}_{\backslash i} }\sum_{a_i}\ket{a_i a_i}^{A_O^{(i)} F_i}\ket{\bm{x}_{\backslash i},\bm{a}_{\backslash i}}^{\bm{A}_{IO}^{(\backslash i)}}\ket{x_i}^{A_I^{(i)}}
\end{align}
with  $\bm{x}_{\backslash i}$ and $\bm{a}_{\backslash i}$ denoting the inputs and outputs of all parties except $i$. We identify the global future of party $i$ opened after purification as the global future of the process, $\HS^{F_i}\equiv\HS^{F}$.
Consider that party $i$ admits auxiliary spaces
$\HS^{A^{(i)}}=\HS^{A^{(i)}_{(I)}}\otimes\HS^{A^{(i)}_{(O)}}$ with $dim(\HS^{A^{(i)}_{(O)}})=dim(\HS^{A^{(i)}_{(I)}})=2$, such that $\HS^{A^{(i)}_{(I)}}\equiv\HS^{P}$ and $\HS^{A^{(i)}_{(O)}}\equiv\HS^{F_t}$, and such that party $i$ performs a SWAP operation between its four systems: the global past wire in $\HS^{P}$ is pulled in the process input space $\HS^{A^{(i)}_O}$, while the process output wire $\HS^{A^{(i)}_I}$ is pulled to the global future $\HS^{F_t}$ (see Fig.~\ref{fig:lugqs} for a depiction of the procedure). 
We then obtain the process matrix of a QC-QC:
  \begin{minipage}{\columnwidth}
  \begin{align}\label{eq:wprev}
    \makebox[\columnwidth][l]{%
      \resizebox{\columnwidth}{!}{%
        $\begin{aligned}
    &W_{QCQC}\\
    &= \ketbra{00}{00}^{PF}\otimes\sum_{\bm{a}_{\backslash i}}\ketbra{\bm{a}_{\backslash i}}{\bm{a}_{\backslash i}}^{\bm{A}_O^{(\backslash i)}}\otimes\ketbra{\bm{x}_{\backslash i}^{a_i=0},x_i}{ \bm{x}_{\backslash i}^{a_i=0},x_i}^{\bm{A}_I^{(\backslash i)}F_t}\\ 
    &+\ketbra{11}{11}^{PF}\otimes\sum_{\bm{a}_{\backslash i}}\ketbra{\bm{a}_{\backslash i}}{\bm{a}_{\backslash i}}^{\bm{A}_O^{(\backslash i)}}\otimes\ketbra{\bm{x}_{\backslash i}^{a_i=1},x_i}{ \bm{x}_{\backslash i}^{a_i=1},x_i}^{\bm{A}_I^{(\backslash i)}F_t}\\ 
    &+\ketbra{00}{11}^{PF}\otimes\sum_{\bm{a}_{\backslash i}}\ketbra{\bm{a}_{\backslash i}}{\bm{a}_{\backslash i}}^{\bm{A}_O^{(\backslash i)}}\otimes\ketbra{\bm{x}_{\backslash i}^{a_i=0},x_i}{ \bm{x}_{\backslash i}^{a_i=1},x_i}^{\bm{A}_I^{(\backslash i)}F_t}\\ 
    &+\ketbra{11}{00}^{PF}\otimes\sum_{\bm{a}_{\backslash i}}\ketbra{\bm{a}_{\backslash i}}{\bm{a}_{\backslash i}}^{\bm{A}_O^{(\backslash i)}}\otimes\ketbra{\bm{x}_{\backslash i}^{a_i=1},x_i}{ \bm{x}_{\backslash i}^{a_i=0},x_i}^{\bm{A}_I^{(\backslash i)}F_t}
    \end{aligned}$%
      }%
    }
  \end{align}
\end{minipage}
with $x_{j}^{a_i=0,1}=w^{a_i=0,1}(\bm{a}_{\backslash \{i,j\}})$. It is straightforward to see that this QC-QC allows to implement the NLWE measurement $(E_{\bm{a}}^{\bm A}=H^{w(\bm{a})}\ketbra{\bm{a}}{\bm{a}}^{\bm A}(H^{w(\bm{a})})^\dagger)$ implementable with the $w$, up to a relabeling $\bm A \rightarrow P\bm A^{({\backslash i})}$, $\bm{a}\rightarrow f,\bm{a}_{\backslash i}$, with the local operations
\begin{minipage}{\columnwidth}
  \begin{align}
    \makebox[\columnwidth][l]{%
      \resizebox{\columnwidth}{!}{%
        $\begin{aligned}
          & \forall j\neq i, \\ \notag
          & M_{a_j}^{A^{(j)}A_{IO}^{(j)}}=\sum_{x_j} H^{x_j}\ketbra{a_j}{a_j}^{A^{(j)}}(H^{x_j})^\dagger\otimes\ketbra{x_j}{x_j}^{A^{(j)}_{I}}\otimes\ketbra{a_j}{a_j}^{A^{(j)}_{O}} \\
          & M_{f}^{FF_t}=\sum_{x_i}H^{x_i}\ketbra{f}{f}^F(H^{x_i})^\dagger \otimes \ketbra{x_i}{x_i}^{F_t}
        \end{aligned}$%
      }%
    }
  \end{align}
\end{minipage}

When $x_i=0$, Fiona measures the control system in $\HS^{F}$ in the computational basis, the causal interferences vanish, projecting $W_{QCQC}$ (Eq.~\eqref{eq:wprev}) on one of its first two terms, depending on the value of $f$. When $x_i=1$, all four terms of $W_{QCQC}$ remain. However, from the transparent control channel condition \ref{eq:transparentb}, $w^{f=0}(\bm{a}_{\backslash i})=w^{f=1}(\bm{a}_{\backslash i})$, i.e. $\bm{x}_{\backslash i}^{a_i=0}=\bm{x}_{\backslash i}^{a_i=1}$. In this case, the target and control communications become independent from one another, and the latter corresponds to an identity channel $\dketbra{\id}{\id}^{PF}$. Hence, apart from the control systems in $\HS^{PF}$, all the communications can be treated classically. Thus, we can write the NLWE measurement in the following LOSupCC form:
        \begin{align}\notag 
    E_{f,\bm{a}_{\backslash i}}^{P\bm A^{({\backslash i})}}&=\delta_{w_i(\bm{a}_{\backslash i}),x_i}H^{x_i}\ketbra{f}{f}(H^{x_i})^\dagger \\ 
    &\bigotimes_{j=1}^{N-1}\delta_{i,j\oplus1}\delta
    _{x_j,w_j^f(\bm{a}_{\backslash j})}M_{a_j|x_j}^{A^{(j)}}.
    \end{align}

\section{From 4-partite generalizations of the Lugano process to QC-QCs}
\label{app:4partite}
As discussed in the main text, whether every process function without a global past satisfies the transparent control channel condition \eqref{eq:transparent} remains an open question. However, we studied several multipartite generalizations of the Lugano process \cite{araujo3,baumeler22,tobar}, each exhibiting a different causal structure while still satisfying this condition. These examples suggest that the answer to this question might be positive.

The first generalization is the Araujo-Guérin-Baumeler process \cite{araujo3}, which we study in the four-partite case but can be extended to an arbitrary number of parties. This process is a direct extension of the Lugano process, as evident from its definition
\begin{align} &x:= d(b\oplus 1)(c\oplus 1) \notag\\ &y:=a(c\oplus 1)(d\oplus 1) \notag \\ &z:=b(d\oplus 1)(a\oplus 1) \notag\\  &w:=c(a\oplus 1)(b\oplus 1). \end{align}
The associated NLWE basis is given by
\begin{align}
        \{\ket{0000},\ket{0101},\ket{0111},\ket{01+0},\ket{01-0},\ket{001+},\ket{001-},\notag\\\ket{1010},\ket{1011},\ket{1101},\ket{1110},\ket{1111},\ket{1+00},\ket{1-00}\notag\\\ket{+001},\ket{-001}\}. \label{eq:4lugset}
    \end{align}
It can be verified that each of the four (or $N$ in the general case) parties satisfies the transparency condition, making it possible to measure the NLWE basis using LOSupCC. However, this process presents a ``drawback'' when generalized to a larger number of parties: as $N$ increases, the winning probability with causal strategies
$P(\bm{x}=w(\bm{a}))$ approaches 1 \cite{baumeler22}, making it increasingly difficult to detect the noncausality of the process.
 
Another $N-$partite generalization of the Lugano process is the Ardehali-Svetlichny process \cite{baumeler22}, which we also studied in the four-partite case:
\begin{align}
    &x:=bc\oplus cd\oplus bd\oplus c,\notag\\  &y:=ac\oplus cd\oplus ad\oplus a\oplus d, \notag\\    &z:=ab\oplus bd\oplus ad\oplus b,\notag\\   &w:= ab\oplus bc\oplus ac\oplus a\oplus c  .
\end{align}
The corresponding NLWE basis is
\begin{align}
        \{\ket{0000},\ket{0+01},\ket{+01+},\ket{001-},\ket{01+0},\ket{+-01},\notag\\\ket{01-0},\ket{0111},\ket{1+0+},\ket{1++-},\ket{-01+},\ket{1+--},\notag\\\ket{1-00},\ket{--01},\ket{111+},\ket{1-1-}\}.
    \end{align}
While this process does not suffer from the same issue as the Araujo-Guérin-Baumeler process when generalized to large $N$, it lacks its full symmetry. Consequently, the transparency condition is not satisfied by all parties but only by some. However, identifying at least one party that meets this condition is sufficient to enable measurement of the NLWE basis using LOSupCC. In this case, only Alice and Charlie satisfy the transparency condition, meaning that only they can serve as the controlling party.

Finally, we also considered the Tobar-Costa process \cite{tobar}, defined as
\begin{align}
    &x:=b(c\oplus d),\notag\\  &y:=c(d(a\oplus 1) \oplus 1), \notag\\    &z:=d(a\oplus 1)(b \oplus 1),\notag\\   &w:= a(b\oplus 1)(c\oplus 1),  
\end{align} 
and the NLWE basis associated with it
      \begin{align}
        \{\ket{0000},\ket{00+1},\ket{0+10},\ket{00-1},\ket{0100},\ket{0111},\notag\\\ket{100+},\ket{100-},\ket{1+10},\ket{1+11},\ket{1100},\ket{1-11},\notag\\\ket{+101},\ket{+-10},\ket{-101},\ket{--10}\}.
    \end{align}
Like the Ardehali-Svetlichny process, this process is not fully symmetric. Consequently, only Alice, Charlie, and Daisy satisfy the transparency condition, making them the only possible control parties.

\section{Causal nonseparability of the SHIFT measurement}
\label{app:cnsshift}
In \cite{dourdent21}, one of us has defined the notion of causal (non)separability for distributed measurements in bipartite and $(2+F)$-partite (two parties and a final party, Fiona, in the causal future of the other two) scenarii, and has shown that that the quantum switch can display a form of noncausality in such semi-device-independent with trusted quantum inputs (SDI-QI) scenario by generating a causally nonseparable distributed measurement (D-POVM), but without the intuitive physical interpretation and relation to NLWE presented here. In (\cite{dourdent24}, Appendix E), the definition and certification were extended to the $(P+2+F)$-partite scenario
by adding a new party, Phil, in the causal past of all other parties. A tripartite notion of causally (non)separable D-POVM was proposed as well in \cite{dourdent22}. In this section, we recall these definitions and clarify how the SHIFT measurement is causally non-separable in both scenarii.

\subsection{$(P+2+F)$-partite scenario (QC-QC-SHIFT)}
We examine the $(P+2+F)$-partite scenario introduced in the main text, which generalizes the previously studied $(2+F)-$partite scenario by introducing an additional party, Phil, who exists in the causal past of all other parties. Phil interacts with the process matrix solely through his output Hilbert space $\HS^P$ without possessing an input Hilbert space. We begin by recalling the definition of causal nonseparability of process matrices in this extended framework before presenting the definition of causal nonseparability of distributed-measurements in which Phil, Alice and Bob act on trusted quantum inputs. 

\subsubsection{Causally (non)separable processes}
Given that Phil is in the global past of the process, and that Fiona is in the global future, the only relevant orders in this scenario are $P\prec A_{IO}\prec B_{IO}\prec F$ and $P\prec B_{IO}\prec A_{IO}\prec F$.
The causally separable process matrices are those that can be decomposed as~\cite{wechs}
\begin{align}
W^{PA_{IO}B_{IO}F} & = W^{P\prec A_{IO}\prec B_{IO}\prec F} + W^{P\prec B_{IO}\prec A_{IO}\prec F} \label{eq:csep_PABF}
\end{align}
such that
\begin{align}
    &\Tr_F W^{P\prec A_{IO}\prec B_{IO}\prec F} = W^{P\prec A_{IO}\prec B_I} \otimes \id^{B_O},\notag\\ & \Tr_F W^{P\prec B_{IO}\prec A_{IO}\prec F} = W^{P\prec B_{IO}\prec A_I} \otimes \id^{A_O}, \notag \\
   & \Tr_{B_I} W^{P\prec A_{IO}\prec B_I} = W^{P\prec A_I} \otimes \id^{A_O},\notag\\ &  \Tr_{A_I} W^{P\prec B\prec A_I} = W^{P\prec B_I} \otimes \id^{B_O}, \notag \\
   & \Tr_{A_I} W^{P\prec A_I} + \Tr_{B_I}W^{P\prec B_I} = \id^{P}, \label{eq:csep_PABF_decomp}
\end{align}
with positive semidefinite matrices $W^{P\prec A_{IO}\prec B_I}\in\L(\HS^{PA_{IO}B_I})$, $W^{P\prec B_{IO}\prec A_I}\in\L(\HS^{PB_{IO}A_I})$, $W^{P\prec A_I}\in\L(\HS^{PA_I})$, $W^{P\prec B_I}\in\L(\HS^{PB_I})$.

In general, the matrices \( W^{P\prec A_{IO}\prec B_{IO}\prec F} \) and \( W^{P\prec B_{IO}\prec A_{IO}\prec F} \) are not necessarily valid (subnormalized) process matrices themselves, unless \( \Tr_{A_I} W^{P\prec A_I} \propto \mathbb{I}^P \) and \( \Tr_{B_I} W^{P\prec B_I} \propto \mathbb{I}^P \). Indeed, causally separable processes allow the ordering between Alice and Bob to be influenced by Phil’s actions in a classically controlled manner~\cite{wechs1}, rather than being restricted to convex combinations of processes with fixed orders \( P\prec A_{IO}\prec B_{IO}\prec F \) and \( P\prec B_{IO}\prec A_{IO}\prec F \). Consequently, such causally separable processes with dynamical causal order cannot be expressed as a convex mixture of valid processes with fixed causal orders.

\subsubsection{Causally (non)separable measurements}
We consider here a $(P+2+F)$-partite scenario in which only Phil, Alice and Bob are provided with quantum inputs $\rho_{\gamma,\alpha,\beta}^{PAB}\in\L(\HS^{PAB})$  (see \cite{dourdent24} for a more general case), and in which Phil only performs a quantum channel from (with abuse of notation) $\L(\HS^{P})$ to $\L(\HS^{P})$, with Choi representation $M^P$. The correlations established by the parties are 
\begin{align}
    &P(f,a,b|\rho_{\gamma,\alpha,\beta}^{PAB})\notag\\&=(M^P\otimes M_a^{AA_{IO}}\otimes M_b^{BB_{IO}}\otimes M_f^{FF_t})*(\rho_{\gamma,\alpha,\beta}^{PAB}\otimes W)\notag\\
    &=E_{f,a,b}^{PAB}*\rho_{\gamma,\alpha,\beta}^{ABC},
\end{align}
where we introduced the D-POVM elements
\begin{align}
E_{f,a,b}^{PAB} = \big(M^{P}\otimes M_a^{AA_{IO}}\otimes M_b^{BB_{IO}}\otimes M_f^{FF_t}\big)*W.
\end{align}
Assuming the process matrix $W\in\L(\HS^{PA_{IO}B_{IO}FF_t})$ is causally separable,  satisfying Eqs.~\eqref{eq:csep_PABF}--\eqref{eq:csep_PABF_decomp}, the induced D-POVM $(E_{f,a,b}^{PAB})_{f,a,b}$ also decomposes as
\begin{align}
    E_{f,a,b}^{PAB} = E_{f,a,b}^{P\prec A\prec B} + E_{f,a,b}^{P\prec B\prec A}\label{eq:Epabf_csep1}
\end{align}
with $(E_{f,a,b}^{P\prec A\prec B})_{f,a,b}$ and $(E_{f,a,b}^{P\prec B\prec A})_{f,a,b}$ such that
\begin{align}
    &E_{f,a,b}^{P\prec A\prec B} \notag\\
    & = (M^{P}\otimes M_a^{A A_{IO}}\otimes M_b^{B B_{IO}}\otimes M_f^{F F_t})* W^{P\prec A_{IO}\prec B_{IO}\prec F}, \notag \\
    \notag \\
    &\sum_f E_{f,a,b}^{P\prec A\prec B} \notag \\ 
    &= (M^{P}\otimes M_a^{A A_{IO}}\otimes \Tr_{B_O}M_b^{B B_{IO}})*W^{P\prec A_{IO}\prec B_I} \notag \\
    & = E_{a,b}^{P\prec A\prec B}, \notag \\ 
    \notag \\
    &\sum_b E_{a,b}^{P\prec A\prec B}  \notag \\
    & = (M^{P}\otimes \Tr_{A_O}M_a^{A A_{IO}})*W^{P\prec A_I} \otimes \id^{B} = E_{a}^{P\prec A} \otimes \id^{B}, \notag \\
    \notag \\
   & \sum_a E_{a}^{P\prec A}  = M^{ P}*\Tr_{A_I}W^{P\prec A_I} \otimes \id^{A} = E^{P\,[A\prec B]} \otimes \id^{A}, \label{eq:Epabf_csep2}
\end{align}
analogously for $E_{f,a,b}^{P\prec B\prec A}$ and  the order $P\prec B_{IO}\prec A_{IO}\prec F$, and with 
\begin{align}
    E^{P\,[A\prec B]} + E^{P\,[B\prec A]} & = M^{P}*\id^P = \Tr_P M^{P}, \label{eq:Epabf_csep3}
\end{align}
such that, since $M^{P}$ is a CPTP map,
\begin{align}
    E^{P\,[A\prec B]} + E^{P\,[B\prec A]} & = \id^{P}. \label{eq:Epabf_csep4}
\end{align}

Note that the individual sets of positive semidefinite matrices $(E_{f,a,b}^{P\prec A\prec B})_{f,a,b}$, $(E_{f,a,b}^{P\prec B\prec A})_{f,a,b}$, $(E_{a,b}^{P\prec A\prec B})_{a,b}$, $(E_{a,b}^{ P\prec B\prec A})_{a,b}$, $(E_{a}^{P\prec A})_{a}$, $(E_{b}^{P\prec B})_{b}$ are not D-POVMs as they do not sum to the identity in general. On the other hand, the set $(E_{f,a,b}^{P\prec A\prec B} ; E_{f,a,b}^{P\prec B\prec A})_{f,a,b}$ formed by all elements of $(E_{f,a,b}^{P\prec A\prec B})_{f,a,b}$ and $(E_{f,a,b}^{P\prec B\prec A})_{f,a,b}$ taken together forms a valid POVM.

\begin{definition}\label{def:csep_DPOVM_P2F}
A ($P$+$2$+$F$)-partite D-POVM $(E_{f,a,b}^{PAB})_{f,a,b}$ that can be decomposed in the form
\begin{align}
E_{f,a,b}^{PAB} = E_{f,a,b}^{P\prec A\prec B} + E_{f,a,b}^{P\prec B\prec A},
\end{align}
where $(E_{f,a,b}^{P\prec A\prec B})_{f,a,b}$ and $(E_{f,a,b}^{P\prec B\prec A})_{f,a,b}$ satisfy
\begin{align}
   \sum_f E_{f,a,b}^{P\prec A\prec B} = E_{a,b}^{P\prec A\prec B}, \quad 
    \sum_b E_{a,b}^{P\prec A\prec B} = E_{a}^{ P\prec A} \otimes \id^{B}, \notag\\
    \sum_a E_{a}^{P\prec A} = E^{P\,[A\prec B]} \otimes \id^{A}, \notag \\
    \sum_f E_{f,a,b}^{P\prec B\prec A} = E_{a,b}^{P\prec B\prec A}, \quad 
    \sum_a E_{a,b}^{P\prec B\prec A} = E_{b}^{P\prec B} \otimes \id^{A}, \notag\\
    \sum_b E_{b}^{P\prec B} = E^{P\,[B\prec A]} \otimes \id^{B}, \quad  E^{P\,[A\prec B]} + E^{P\,[B\prec A]} = \id^{P}
\end{align}
(with all operators $E_{\cdots}^{\cdots} \ge 0$) is said to be \emph{causally separable}.
\end{definition}
A causally separable ($P$+$2$+$F$)-partite process matrix can only generate a causally separable ($P$+$2$+$F$)-partite D-POVM. \emph{A contrario}, if one finds that, for some choice of operations for each party, the induced D-POVM is causally nonseparable, this certifies in a SDI-QI manner that the process matrix that generated it is itself causally nonseparable.
This is the case for the quantum switch (Eq.~\eqref{eq:qs}) and the QC-QC (Eq.~\eqref{eq:qcqc}), which, with the operations Eq.~\eqref{eq:qlocop} /\eqref{eq:qclop}, induce the SHIFT measurement. Its causal nonseparability in the ($P$+$2$+$F$)-partite scenario can thus be certified by  providing the states of the SHIFT basis, which serves as a causal witness \cite{araujo1,branciard,dourdent21,dourdent24}, to the parties, and verify that the witness inequality 
\begin{equation}\label{eq:p2fineq}
\resizebox{\columnwidth}{!}{$\displaystyle
     {
\frac{1}{8}}\sum_{\gamma,\alpha,\beta} P(f=\gamma,a=\alpha,b=\beta |\ket{\psi}_{\gamma,\alpha,\beta}^{PAB}=H^{w_{L}(\gamma,\alpha,\beta)}\ket{\gamma,\alpha,\beta} ) \leq \xi$}
\end{equation}
is violated, with  $\xi$ denoting the causal bound. In fact, since the SHIFT measurement is linked to the noncausal Lugano process, we can conclude that it is causally nonseparable, regardless of how it is constructed. Since causally separable processes can only generate causally separable distributed measurements,  $\xi<1$ (the algebraic maximum). Using a semi-definite program (SDP) and a see-saw algorithm (similar to the approach in \cite{branciard1}), optimizing on the process matrix as well as the local operations, we find a numerical causal bound of $\approx 0,9268$. It is straightforward that the witness inequality Eq.~\eqref{eq:p2fineq} is maximally violated by the SHIFT measurement. 
Notably, a very similar causally non-separable D-POVM is certified in (\cite{dourdent24}, Appendix E), with a quantum switch and similar local operations, but with quantum inputs only for Phil (Alice and Bob receive classical inputs). 

\subsection{Tripartite scenario (Lugano-SHIFT)}
A definition of causal (non-)separability for distributed measurements in a tripartite scenario was proposed in \cite{dourdent22}. We begin by recalling the definition of causal nonseparability of tripartite process matrices, before introducing the definition for D-POVM and explaining how the SHIFT measurement is (trivially) causally nonseparable in this scenario.

\subsubsection{Causally (non)separable processes}
From \cite{wechs}, a tripartite process matrix is causally separable if and only if it can be decomposed as 
\begin{align}
   & W^{sep} = W_{(A)}+W_{(B)}+W_{(C)},\notag\\
    &W_{(A)}=W_{(ABC)}+W_{(ACB)},\notag\\
    &W_{(B)}=W_{(BCA)}+_{(BAC)},\notag\\
    &W_{(C)}=W_{(CAB)}+W_{(CBA)}
    \label{eq:cstri}
\end{align}
where, for each permutation of the three parties $(X,Y,Z)$, $W_{(X,Y,Z)}$ and $W_{(X)}:=W_{(X,Y,Z)}+W_{(X,Z,Y)}$ are positive semidefinite matrices satisfying 
\begin{align}
    _{[1-X_O]YZ}W_{(X)}&=0 \notag\\
    _{[1-Y_O]Z}W_{(X,Y,Z)}&=0,\notag\\ _{[1-Z_O]}W_{(X,Y,Z)}&=0
    \label{eq:cstri1}
\end{align}
with the trace-and-replace notation:

$_XW=\frac{\id^{X}}{d_X}\otimes \Tr_X W, \hspace{3mm} _{[1-X]}W=W - _XW$.

$W^{sep}$ is decomposed in terms of three nonnormalized process matrices $W_{(X)}$, which can be interpreted as a process compatible with ``the party $X$ acts first''. Note also that the $W_{(X,Y,Z)}$ are not necessarily valid process matrices.  Thus, a causally nonseparable tripartite process matrix is the convex mixture of the three valid process matrices each associated with a different first party.

\subsubsection{Causally (non)separable measurements}
Consider the general tripartite scenario involving Alice, Bob and Charlie, and consider the non-valid process matrix $W_{(ABC)} = W_{(A B C_I)}\otimes \id^{C_O}$ of Eq.~\eqref{eq:cstri}-\eqref{eq:cstri1}. Then one has
\begin{align}
&\sum_c E_{a,b,c}^{ABC} \notag\\&=  \sum_c (M_a^{AA_{IO}}\otimes M_b^{BB_{IO}}\otimes M_c^{CC_{IO}})*(W_{(A B C_I)}\otimes \id^{C_O}) \notag \\
& = (M_a^{AA_{IO}}\otimes M_b^{BB_{IO}}\otimes \Tr_{C_O}  \sum_c \,M_c^{CC_{IO}})* W_{(A B C_I)} \notag \\
& = (M_a^{AA_{IO}}\otimes M_b^{BB_{IO}}\otimes \id^{CC_I})* W_{(A B C_I)} = E_{a,b}^{AB} \otimes \id^{C}
\label{eq:dpovmtri}
\end{align}
with $E_{a,b}^{AB} = (M_a^{AA_{IO}}\otimes M_{b}^{BB_{IO}}) * \Tr_{C_I} W_{(A_{IO} B_{IO} C_I)} \ge 0$.\\

Moreover, using 
\begin{equation}
    \Tr_{C_I}W_{(ABC_I)}=\frac{1}{d_{B_O}}\Tr_{B_O C_I}W_{(A B C_I)}\otimes\id^{B_O}
\end{equation} we also obtain
\begin{align}
\sum_{b,c} E_{a,b,c}^{ABC} & = \sum_b E_{a,b}^{AB} \otimes \id^{C}=E_a^{A}\otimes\id^{BC} 
\label{eq:dpovmtri1}
\end{align}
with $E_a^{A}=\frac{1}{d_{B_O}}\Tr_{B C_I} W_{(A B C_I)}*M_a^{AA_{IO}}\geq 0$.  Eq.~\eqref{eq:dpovmtri}-\eqref{eq:dpovmtri1} can be interpreted as no-signalling conditions \cite{hoban18}, meaning respectively that Charlie cannot signal to Bob nor Alice, and Bob cannot signal to Alice. Note however that because $W_{(ABC)}$ is not a valid process matrix in general (cf. Eq.~\eqref{eq:cstri}-\eqref{eq:cstri1}), $(E_a^{A})_a$ (and thus $(E_{a,b}^{AB})_{a,b}$ and $(E_{a,b,c}^{ABC})_{a,b,c}$) is not a POVM, $\sum_a E_a^{A}\neq \id^{A}$ in general. \\

We need to consider the complementary $(A,C,B) $ such that, defining \[E_{a,b,c}^{(A)}:=E_{a,b,c}^{(A,B,C)}+E_{a,b,c}^{(A,C,B)},\]  the set $(E_{a,b,c}^{(A)})_{a,b,c}$ is a D-POVM:
\begin{align}
\sum_{a,b,c} E_{a,b,c}^{(A)} & =\sum_a M_a^{AA_{IO}} * (\frac{1}{d_{B_O}}\Tr_{B_{IO} C_I} W^{(A B C_I)}\notag\\
&+\frac{1}{d_{C_O}}\Tr_{C_{IO} B_I} W^{(A C B_I)})\otimes\id^{BC}\notag\\
&=\sum_a M_a^{AA_{IO}} *(\rho^{A_I}\otimes\id^{A_O})\otimes\id^{BC}\notag\\
&=\id^{A_I}*\rho^{A_I}\id^{ABC}=\id^{ABC}
\end{align}
with $\rho^{A_I}$ such that $\frac{1}{d_{B_O}d_{C_O}}\Tr_{B_{IO}C_{IO}}W_{(A)}=\frac{1}{d_{B_O}d_{C_O}}\Tr_{BC}(W_{(ABC)}+W_{(ACB)})=\rho^{A_I}\otimes\id^{A_O}$.\medskip \\
Thus, we can extend the notion of causal (non)separability for bipartite D-POVM to a general tripartite case.\\

\begin{definition} A tripartite D-POVM $(E_{a,b,c}^{ABC})_{a,b,c}$ that can be decomposed as a convex mixture of D-POVMs compatible with ``$A$ first'', ``$B$ first'' and ``$C$ first'', i.e., of the form
\begin{align}
    E_{a,b,c}^{ABC} &= E_{a,b,c}^{(A)}+E_{a,b,c}^{(B)}+E_{a,b,c}^{(C)}
\end{align}
is said to be causally separable.\\
Each D-POVM $(E_{x,y,z}^{(X)})_{x,y,z}$ decomposes as a sum of sets of positive semidefinite matrices (but whose elements do not sum up to $\id^{X}$ in general)  $(E_{x,y,z}^{(X,Y,Z)})_{x,y,z}+(E_{x,y,z}^{(X,Z,Y)})_{x,y,z}$. For each permutation of $(A,B,C)$, the elements $E_{x,y,z}^{(X,Y,Z)}$ and $E_{x,y,z}^{(X)}:=E_{x,y,z}^{(X,Y,Z)}+E_{x,y,z}^{(X,Z,Y)}$ are positive semidefinite matrices satisfying 
\begin{align}
\sum_{x,y,z}E_{x,y,z}^{(X)}&=\id^{XYZ}\notag\\
\sum_{y,z}E_{x,y,z}^{(X,Y,Z)}&=E_x^X \otimes \id^{YZ}\notag\\
\sum_{z}E_{x,y,z}^{(X,Y,Z)}&=E_{x,y}^{(X,Y)}\otimes\id^Z
\end{align}
\end{definition}
Note that the question of whether any tripartite causally separable D-POVM can be obtained from a tripartite causally separable process matrix is an open problem. Nevertheless, a causally separable tripartite process matrix can only generate a causally separable tripartite D-POVM. Thus, if one finds that, for some choice of operations for each party, the induced D-POVM is causally nonseparable, this certifies in a SDI-QI manner that the process matrix that generated it is itself causally nonseparable. 
Clearly, if a process matrix is noncausal (i.e. it violates a causal inequality), it can also be certified in a SDI-QI manner, and thus generate a causally nonseparable D-POVM (see \cite{dourdent21}, Appendix F). It is also clear that the SHIFT measurement, and more generally, any tripartite NLWE measurement induced by a process function without global past, is causally nonseparable in this sense.  

\section{Network-Device-Independent certification of superposition of classical communications ---}\label{app:ndi}
In the $(P+2+F)-$partite scenario, the QC-QC (Eq.~\eqref{eq:qcqc}) can induce the SHIFT measurement, and thus maximally win the game consisting in perfectly discriminating the SHIFT basis states $\{\ket{\psi}_{\gamma,\alpha,\beta}^{PAB}=H^{w_{L}(\gamma,\alpha,\beta)}\ket{\gamma,\alpha,\beta}^{PAB}\}_{\gamma,\alpha,\beta}$, i.e. maximally violate the inequality Eq.~\eqref{eq:p2fineq}. Let us simplify the scenario by leveraging the fact that the input states are product states, and by replacing the quantum inputs from Alice and Bob with black boxes producing $a$ and $b$, respectively. 

\begin{figure}[ht]
	\begin{center}
	\includegraphics[width=0.85\columnwidth]{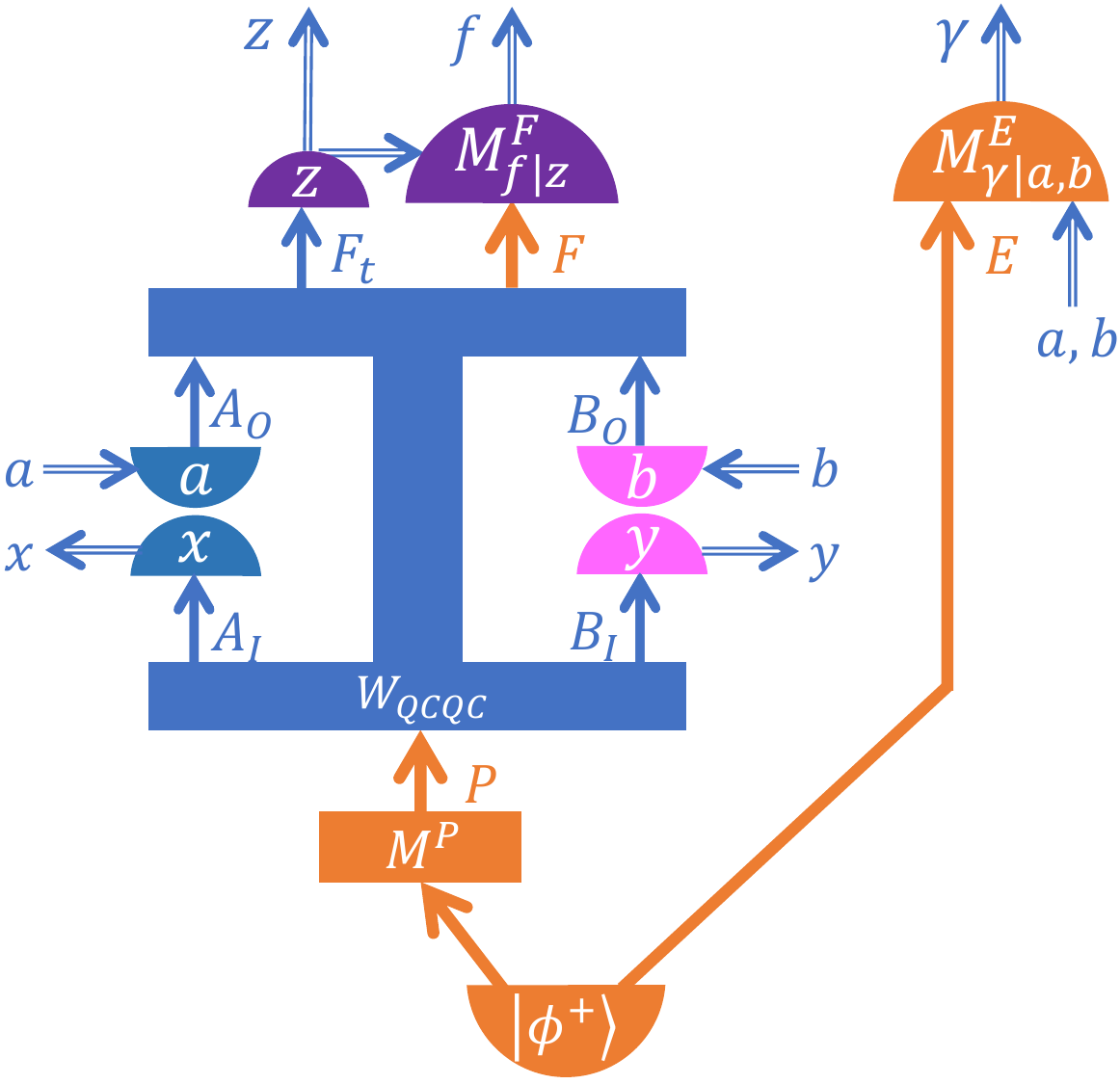}
	\end{center}
	\caption{Towards the NDI certification of superposition of classical communications. The next stage consists in considering Phil and Eve's operations as black boxes with additional classical inputs, such that they can self-test the quantum inputs $\ketbra{\psi}{\psi}^P_{\gamma|a,b}$ that are used to certify the causal nonseparability of the D-POVM $(E_{f,x,y,z|a,b}^P)_{f,x,y,z|a,b}$ induced by the local operations on the process $W_{QCQC}$. } 
	\label{fig:diqs}
\end{figure}

Consider that, instead of SHIFT basis states, Alice and Bob receive computational basis states $\{\ket{a}^A\}_a$ and $\{\ket{b}^B\}_b$ respectively,  and Phil receives BB84 states $\{\ket{\psi}^P_{\gamma|a,b}=H^{b(a\oplus 1)}\ket{\gamma}^P\}_{\gamma|a,b}$, with $a,b,\gamma\in\{0,1\}^3$. We define the quantum operations:
\begin{align}
    M_x^{AA_{IO}}&=\sum_a\ketbra{a}{a}^A\otimes \ketbra{x}{x}^{A_I}\otimes\ketbra{a}{a}^{A_O},\notag\\
    M_y^{BB_{IO}}&=\sum_b\ketbra{b}{b}^B\otimes \ketbra{y}{y}^{B_I}\otimes\ketbra{b}{b}^{B_O},\notag\\
    M_{f,z}^{FF_t}  &=(H^z\ketbra{f}{f}(H^z)^\dagger)^{F}\otimes\ketbra{z}{z}^{F_t}
\end{align}
It is easily verified that $(M_x^{AA_{IO}})_x$ and $(M_y^{BB_{IO}})_y$ are valid quantum instruments ($\Tr_{A_O}\sum_x M_{x}^{AA_{IO}}=\break\sum_a \ketbra{a}{a}^A\sum_x \ketbra{x}{x}^{A_I}=\id^{AA_I}$).
Fiona's operations are unchanged with respect to Eq.~\eqref{eq:qclop}, the only difference being that we consider $z$ as an explicit output here, such that $M_{f,z}^{FF_t}$ can be seen as a causally separable D-POVM where the operation on $F_t$ is in the causal past of the one on $F$. Alice and Bob's operations are such that $x$ and $y$ are not used as inputs resulting from measuring their process input spaces and labeling the measurements on the auxiliary systems, but as genuine outputs. With this choice, the induced D-POVM elements are
\begin{align}
    E_{f,x,y,z}^{PAB}=\sum_{a,b}E_{f,x,y,z|a,b}^P\otimes\ketbra{a}{a}^A\otimes\ketbra{b}{b}^B
\end{align}
with $E_{f,x,y,z|a,b}^P=(\ketbra{x}{x}^{A_I}\otimes\ketbra{a}{a}^{A_O}\otimes\ketbra{y}{y}^{B_I}\otimes\ketbra{b}{b}^{B_O}\otimes M_{f,z}^{FF_t})*W_{QCQC}$, and so that $E_{f,x,y,z|a,b}^P=E_{f,x,y,z}^{PAB}*(\ketbra{a}{a}^A\otimes\ketbra{b}{b}^B)$. We can thus simplify the scenario by considering that Alice and Bob are not provided with auxiliary quantum systems anymore, but simply receive classical inputs $a$ and $b$ respectively. The correlations established by the parties are then
\begin{align}
    P(f,x,y,z|a,b,\ket{\psi}_{\gamma|a,b}^P)=E_{f,x,y,z|a,b}^P*\ketbra{\psi}{\psi}_{\gamma|a,b}^P
\end{align}
Noticing that the structure of the Lugano process is still preserved in the QC-QC, the latter thus maximally violates the inequality 
\begin{equation}
\resizebox{\columnwidth}{!}{$\displaystyle
\frac{1}{8} \sum_{a,b,\gamma} P(f=\gamma,(x,y,z)=w_L(a,b,\gamma)|\ket{\psi}^P_{\gamma|a,b}=H^{b(a\oplus 1)}\ket{\gamma})\leq \xi$}
\end{equation}
This inequality is the SDI-QI analog of the standard DI inequality maximally violated by the Lugano process, $P((x,y,z)=w_{L}(a,b,c))\leq 3/4$. It provides a semi-device-independent certification of the superposition of classical communications in $W_{QCQC}$. This method can be extended to a fully device-independent certification by considering an additional party, Eve, who prepares remotely Phil's quantum inputs by performing suitable operations on her part of a maximally entangled state shared with him, $\ketbra{\psi}{\psi}^P_{\gamma|a,b}\propto M_{\gamma|a,b}^E*\ketbra{\phi^+}{\phi^+}^{PE}$ (Fig.~\ref{fig:diqs}), and by self-testing these quantum inputs, following the network-device-independent (NDI) protocol described in \cite{dourdent24}.\\

\section{Towards an experimental realization of the SHIFT measurement }
\label{app:exp}
This work addresses the physical realizability of the Lugano process—along with its implications for non-causality and the violation of causal inequalities—by demonstrating that one of its key signatures, the implementation of the SHIFT measurement, can be reproduced within a quantum circuit with quantum control of causal orders (QC-QC) in a scenario with quantum inputs. Given that QC-QCs are considered to ``encompass all known examples of physically realizable processes with indefinite causal order'' \cite{wechs1}, our result provides evidence supporting the possibility of a faithful physical simulation of the Lugano process.  

Moreover, several experimental implementations of the quantum switch, primarily in photonic setups \cite{procopio,rubino,goswami18,wei19,rubino22,goswami20, guo20,taddei20, rubino21, cao23,stromberg23,antesberger24} (see \cite{rozema24} for a review), suggest that an experimental realization of the SHIFT measurement—and thus a simulation of the Lugano process—may be within reach. However, the extent to which these experiments genuinely realize indefinite causal orders, rather than merely simulating effects that could only be achieved through superposed gravitational processes, has been a topic of debate \cite{maclean17,oreshkov19,paunkovic20,ormrod23,vilasini24,vilasini24a}.  

After briefly outlining this controversy, we will explore how our transformation of the Lugano process into a QC-QC relates to the framework of time-delocalized subsystems and how a photonic implementation of the SHIFT measurement could be achieved. Finally, we will discuss eventual implications of a gravitational SHIFT measurement for causal indefiniteness.

 \subsection{The causal omelette}

The research program on indefinite causal orders originates from the introduction of an operational framework, the causaloid formalism \cite{hardy05}, which aimed at bridging the gap between the non-fixed causal structure of general relativity and the non-fixed quantities of quantum theory. ``It is therefore likely that, in a theory of quantum gravity, we will have indefinite causal structure.'' \cite{hardy07}. Such structures were later considered as enabling a new form of higher-order computation \cite{hardy09,chiribella13}, leading to the development of another framework, the process matrix formalism \cite{oreshkov1}, which demonstrated that processes with indefinite causal orders are ``allowed by quantum theory'', in the sense that they can arise between separated closed laboratories performing local quantum operations. 

Crucially, the derivation of such processes relies on relaxing the assumption that there exists a spacetime on which the evolution of quantum systems and the constraints given by relativity are defined. The causal relata, the elements involved in causation, i.e. here the ``local events'', correspond to generations of classical outcomes from classically labeled quantum operations performed in separated closed laboratories. These laboratories are defined by the distinct quantum systems they act on, which, importantly, can be decomposed into nontrivial subsystems associated with different times, the so-called ``time-delocalized subsystems'' \cite{oreshkov19, Wechs23}. Causal relations are thus always defined in an operational manner, namely as one-way non-signaling constraints, encapsulating the order in which these operations are performed.\\

Because they map output Hilbert spaces of parties into their input spaces, process matrices are sometimes interpreted as ``having the form of a closed time-like curve (CTC), where information is sent back in time through a noisy channel'' \cite{oreshkov1}. Similarly, process functions have been described as ``reversible time travel or dynamics with CTCs compatible with freedom of choice'' \cite{baumeler19, tobar}, avoiding famous time-travel paradoxes such as the grandfather paradox, where an effect suppresses its own cause, and the information paradox, where an effect is its own cause \cite{baumeler, baumeler21}. In fact, process matrices can be seen as the linear subset of an \textit{operational, information-theoretic} analog of genuine spacetime CTCs, based on post-selected quantum teleportations \cite{araujo3}, the so-called ``post-selected closed timelike curves'' (P-CTCs) \cite{politzer,bennett,svetlichny}. The use of this spacetime terminology aligns more closely with the well-established notions of \textit{events} and \textit{causality} in relativity, i.e., points (or regions) in spacetime and properties of spacetime geometry. Moreover, one may argue that ``in physical experiments, information-theoretic structures are embedded in spacetime, and these two notions of causality play together in a compatible manner.'' \cite{vilasini24}.\\

Depending on the notion of \textit{event} used, the interpretation of experiments implementing indefinite causal orders may vary significantly. On the one hand, time-delocalized operations in a photonic quantum switch are applied once and only once, leading to the conclusion that they do indeed realize indefinite causal orders. Although it has been argued that these may only be realized in a \textit{``weak''}, coarse-grained sense \cite{ormrod23}, they may still provide a sufficient information-based quantifiable resource advantage \cite{fellous23}. On the other hand, adopting a spacetime-based definition of events, photonic experiments—being embedded in a quasi-Minkowski flat spacetime—can only feature indefinite causal orders in a coarse-grained description of the process \cite{vilasini24}, and are thus sometimes designated as ``simulations''. Genuine fine-grained indefinite causal orders would then only be implementable in realizations involving superpositions of gravitational fields, the so-called ``gravitational quantum switch'' \cite{zych,paunkovic20,moller21,moller24}. This fine-graining versus coarse-graining distinction has also been extended to the more general class of QC-QCs \cite{salzger25}. In this context, it has been shown that the causal-box framework—which models composable quantum information protocols embedded in spacetime—corresponds to a fine-grained description of QC-QCs, in which an indefinite causal structure is resolved into a collection of quantum operations arranged in a well-defined, acyclic causal order compatible with the underlying spacetime structure.
In an operational framework, events and their localisation have also been defined relative to a ``Lab'',  by specifying a reference physical degree of freedom \cite{vilasini25}. This reference carries a property called relative measurability, which captures how correlations between the Lab’s reference and the systems under study allow one to localize and order events. Within this framework, the distinction between photonic and gravitational implementations of the quantum switch does not originate from gravity per se, but rather from distinct assumptions on the Labs, more precisely differences in the quantum correlations between the spatiotemporal reference degrees of freedom and the processes being investigated.

Alternatively, other relativistic definitions of event (in terms of coincidences of worldlines) and causal order, encompassing both the photonic and gravitational quantum switch without hierarchy, have been proposed \cite{delahamette24a}. Since fine-grained structures, such as the spacetime location of an event, are operationally inaccessible—as any measurement would destroy causal interferences—one might adopt an attitude of epistemological modesty \cite{grinbaum10} and favor coarse-grained approaches \cite{delahamette24b}.  
Finally, a recent result demonstrates that the quantum switch cannot be simulated by any quantum circuit, even when allowing for a greater number of local (non-unitary) operations \cite{bavaresco24}. This further strengthens the case that photonic implementations offer a genuine and effective way to realize indefinite causal orders.

\subsection{Converting the Lugano process into a
QC-QC via time-delocalised subsystems}

\begin{figure}[ht]
	\begin{center}
	\includegraphics[width=0.98\columnwidth]{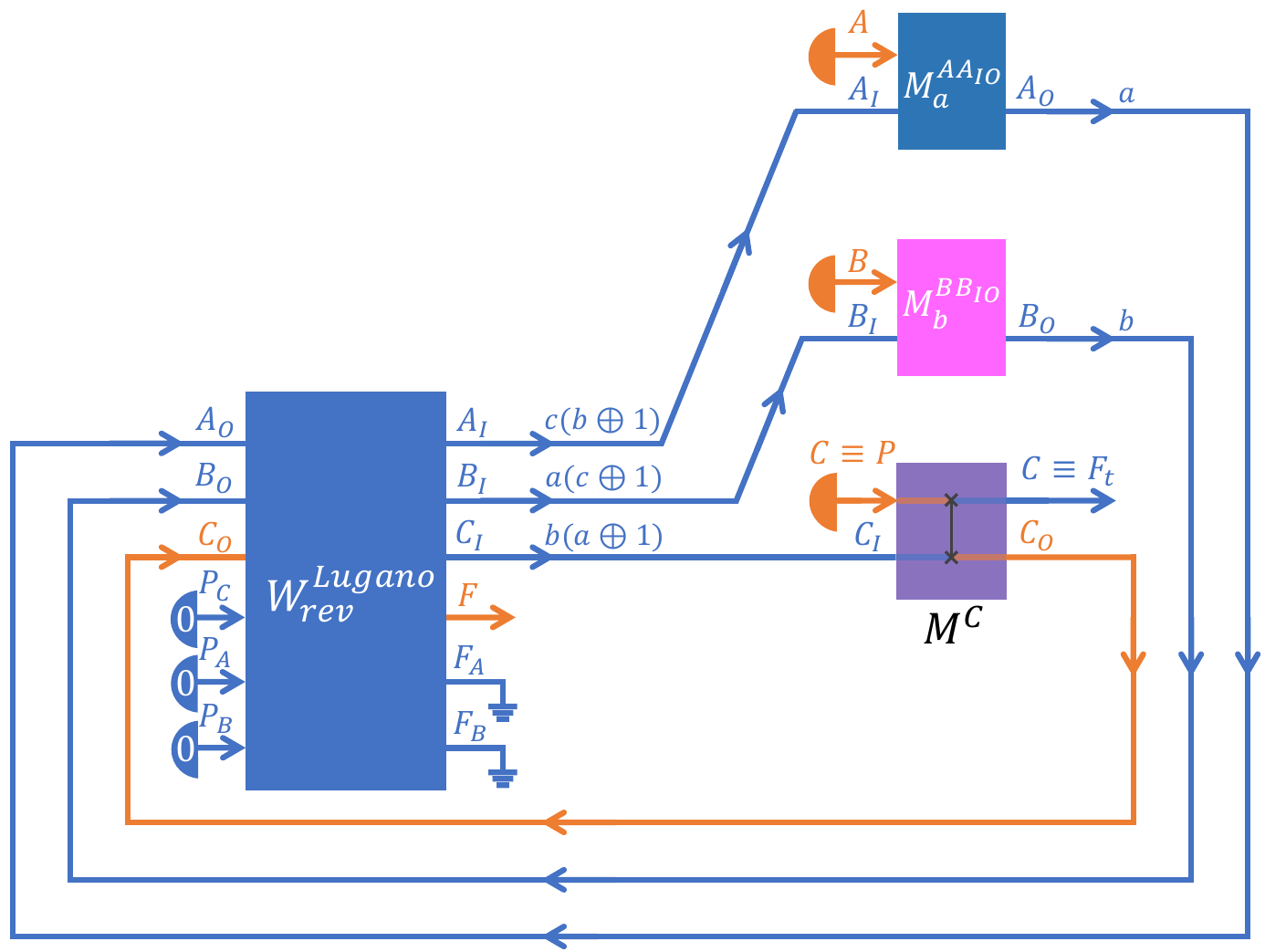}
	\end{center}
	\caption{Circuit representation of the composition of the partially purified Lugano process, Alice and Bob's operations, and Charlie's SWAP operation leading to the QC-QC. The quantum control channel is represented by the orange wire, and classical communications by blue wires (based on Fig.21 in \cite{Wechs23}).}
	\label{fig:lugdeloc}
\end{figure}

\begin{figure*}[htb]
	\begin{center}
	\includegraphics[width=1.7\columnwidth]{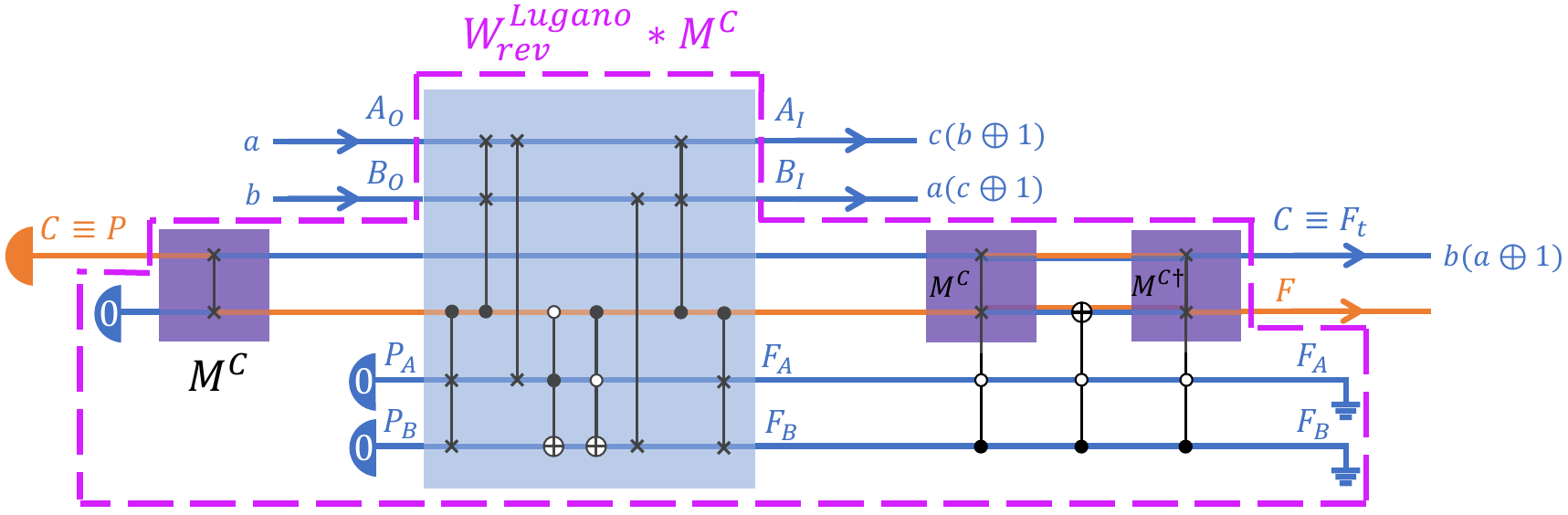}
	\end{center}
	\caption{Circuit representation of the composition of the partially purified Lugano process and Charlie's time-delocalized SWAP operation leading to the QC-QC. Charlie's
  subsystems $C_I$ and $C_O$ are ``time-delocalized'' (based on Fig.22 in \cite{Wechs23}).}
	\label{fig:timedeloc}
\end{figure*}

In \cite{Wechs23}, it was shown that the Lugano process admits a realization on time-delocalized subsystems, i.e. can be interpreted in terms of a quantum circuit with quantum systems that are delocalized in time \cite{oreshkov19}. This realization offers as well an intuitive understanding of our transformation of the Lugano process into a QC-QC by swapping Charlie's systems (Section \ref{app:lugqcqc}, Fig.~\ref{fig:lugdeloc}). In fact, in the circuit representation of Charlie's time-delocalized operation composed with the Lugano process (Fig.~\ref{fig:timedeloc}), one can identify that while Alice and Bob's variables are simply time-delocalized by being classically controlled by Charlie's process systems, the latter are time-delocalized in a different way. The operation $M^C$ is applied in a time-localized manner at the beginning of the circuit and may later be reversed and reapplied at the end, conditioned on the outputs of Alice and Bob. However, as pointed out in \cite{Wechs23}, this should not be mistaken for Charlie's operation being performed multiple times, which would lead to a contradiction with spacetime causality by involving a naive form of retrocausal signalling. Like Alice and Bob's operations, it is executed only once, acting on time-delocalized input and output systems. 

Charlie performing a SWAP operation between his (classical) process and (quantum) auxiliary systems has three key consequences. First, the process becomes transparent to the auxiliary quantum system initially prepared in \( \mathcal{H}^C \equiv \mathcal{H}^P \), which remains unchanged upon reaching \( \mathcal{H}^F \). Second, the control over the order of Alice and Bob’s operations is now governed by this quantum system rather than Charlie’s classical output system. Third, Charlie’s classical input system in $\HS^{C_I}$ is directly transferred to the global future, \( \mathcal{H}^C \equiv \mathcal{H}^{F_t} \).  

As a result, the SWAP operation effectively replaces a time-delocalized classical control with a quantum control of Alice and Bob’s causal order.

\begin{figure}[htb]
	\begin{center}
	\includegraphics[width=0.95\columnwidth]{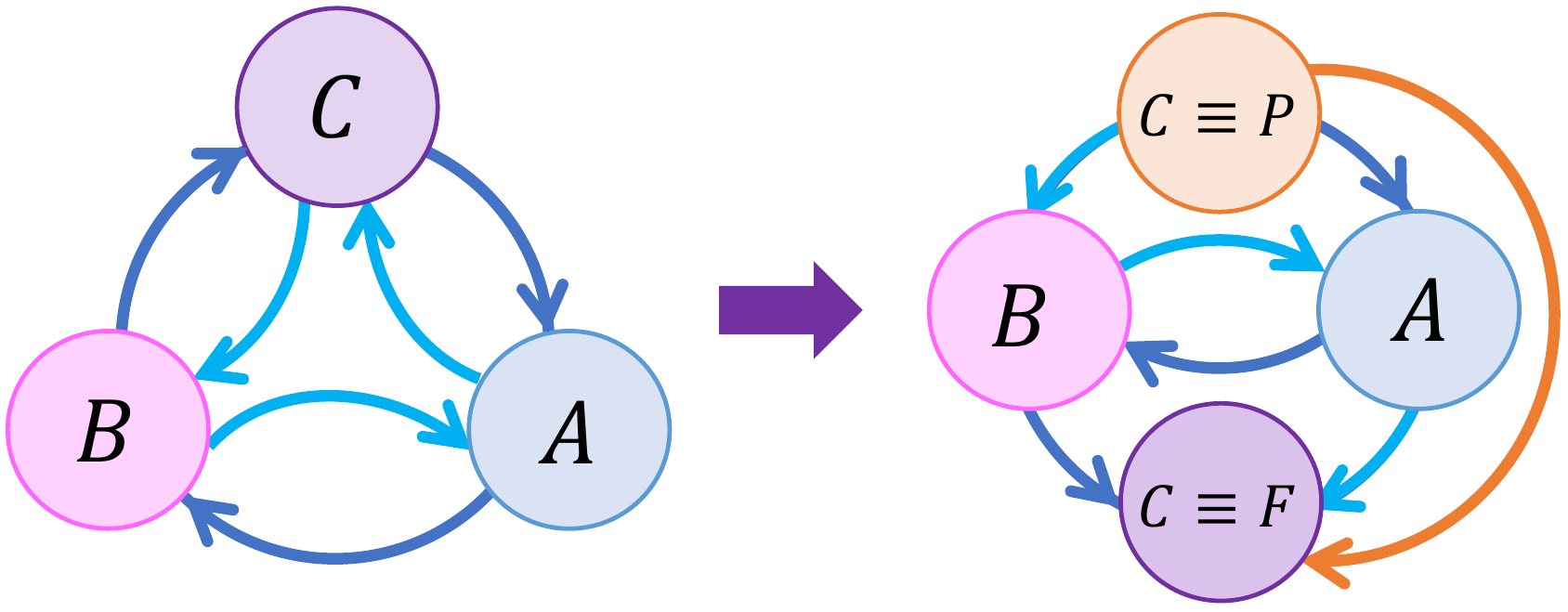}
	\end{center}
	\caption{From the causal structure of the Lugano process (left), to the causal structure of a QC-QC (right).}
\label{fig:causalgraph}

\end{figure}

This suggests that the time-delocalized party, Charlie, can be effectively interpreted as being split into two distinct parties, Phil and Fiona. This perspective is illustrated by analyzing the transformation of the Lugano process into a QC-QC based on their respective causal structures (Fig.~\ref{fig:causalgraph}). By unfolding Charlie's node into separate past and future nodes within the Lugano causal structure, one naturally arrives at a QC-QC structure. 

By only requiring an additional quantum auxiliary system and the transparent control channel condition Eq.~\eqref{eq:transparent}, any $N-$ partite Boolean process function without global past can be converted into a $P+(N-2)+F$ causally nonseparable QC-QC by following a similar method.

\label{app:timedeloc}

\subsection{A photonic SHIFT measurement}

Various photonic quantum switches have been implemented (cf. review \cite{rozema24}). In \cite{wechs1,dourdent22}, one of us has proposed a photonic quantum switch scheme that would 
avoid a redundancy encoding the control system in more than one degree of freedom, which can be found in most of the photonic realizations.

\begin{figure}[htb]
	\begin{center}
	\includegraphics[width=0.98\columnwidth]{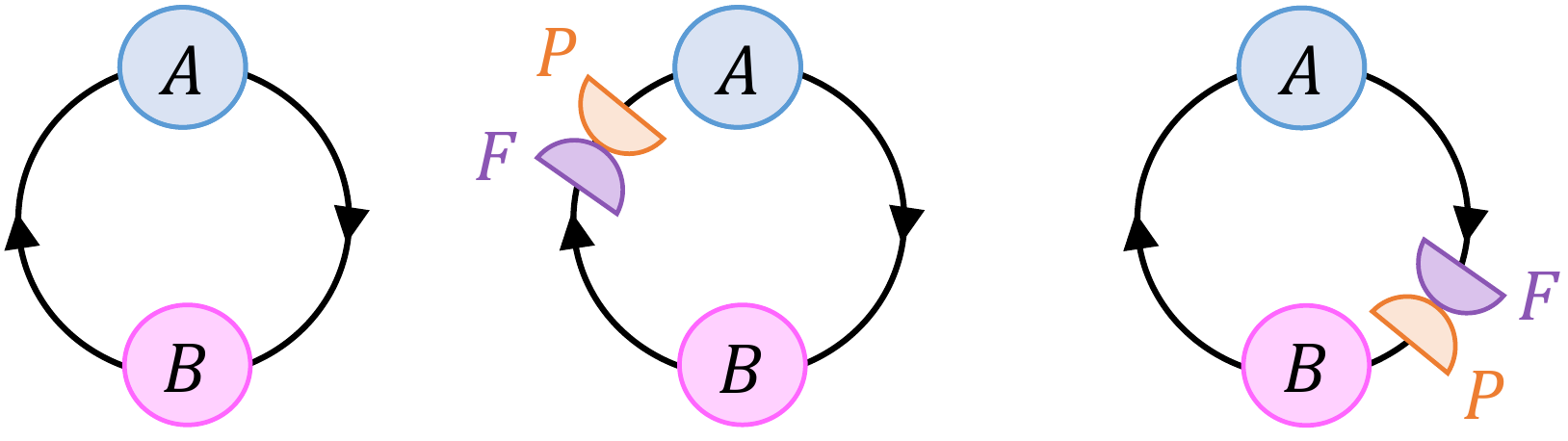}
	\end{center}
	\caption{ Trapped inside a causal loop, Alice and Bob’s operations
cannot be causally ordered (left). The introduction of a cut defines a global past and a global
future, and thus a well-defined causal relation between Alice and Bob. If the cut is put
between Bob’s output space and Alice’s input space, the causal order “Alice is in the
causal past of Bob” is defined (center). If it is put between Alice’s output space and Bob’s input
space, the causal order “Bob is in the causal past of Alice” is defined (right). Making the position
of the cut undetermined leads to an indefinite causal relation. \cite{dourdent22}}
	\label{fig:loopcut}
\end{figure}

The underlying concept of this alternative implementation aligns with the following intuition on indefinite causal orders, presented in \cite{dourdent22}. Suppose Alice and Bob’s laboratories are enclosed within a closed global causal loop, where Alice’s output directly feeds into Bob’s input and vice versa, effectively trapping the exchanged information. Within such a loop, conventional notions of causality and time lose their meaning. To establish a well-defined causal relation between Alice and Bob, the loop must be ``cut." This cut serves as a fundamental primitive, from which the notion of causal relation emerges by defining an entry point (the global past $P$) and an exit point (the global future $F$). The precise location of the cut determines the causal relationship between Alice and Bob’s operations, as illustrated in Fig.~\ref{fig:loopcut}. If the position of the cut remains undetermined, the causal order of Alice and Bob’s operations becomes indefinite.

In a photonic circuit, this cut could be physically implemented using fast-switching removable mirrors \cite{wechs1}. These mirrors would be temporarily removed between the application of Alice and Bob’s operations, allowing both operations to be performed while preserving the possibility of an indefinite causal order. A similar concept was independently realized in \cite{antesberger24}, where active photonic elements were used to deterministically generate and manipulate time-bin encoded qubits.

\begin{figure}[hb]
	\begin{center}
	\includegraphics[width=0.98\columnwidth]{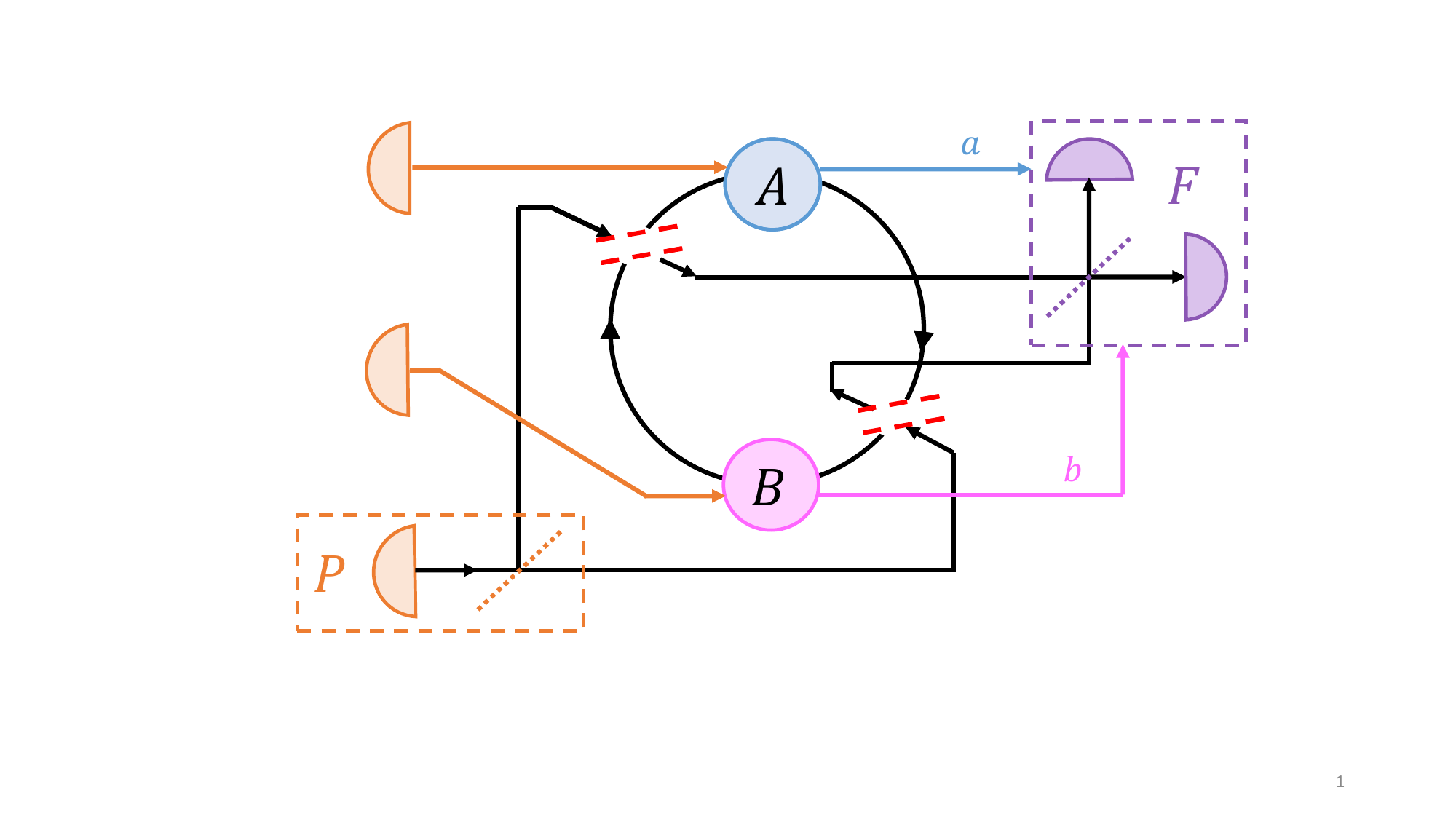}
	\end{center}
	\caption{ A possible scheme for an implementation of the SHIFT measurement via a photonic quantum switch, in which
the control qubit gets encoded in the path degree of freedom, and the target system in an
internal degree of freedom of the photon. The dashed red photonic elements represent reflecting mirrors that are temporarily removed between the applications of Alice and Bob’s operations, allowing them to be performed in either order. Together with the superposed controlling path, they act as the undetermined cut of Fig.~\ref{fig:loopcut}. To realize the SHIFT measurement, Alice and Bob would receive additional photons, where the SHIFT state is encoded, for example, in their polarization. They would then communicate their respective outcomes, $a$ and $b$, to Fiona, who would determine whether to use the photonic circuit in a which-path configuration (with detectors at the end of each path) or in an interferometric configuration (with a beam splitter and detectors) based on the value of $b(a\oplus 1)$.}
	\label{fig:shiftexp}
\end{figure}

This photonic quantum switch could be interpreted as a ``causal loop with an undetermined cut" implemented within a Mach–Zehnder interferometer. Introducing auxiliary photons, such a setup
could enable a photonic realization of the SHIFT measurement, ultimately allowing for a photonic simulation of the Lugano process (see Fig.~\ref{fig:shiftexp}).\\

The main challenge in implementing a photonic SHIFT measurement will be the realization of Alice and Bob’s causally ordered measure-and-reprepare operations. Although several photonic quantum switches have been demonstrated with measure-and-reprepare operations \cite{rubino, cao23, antesberger24}, no existing implementation enforces a measurement on the process input that causally influences a subsequent measurement on the auxiliary system, with the outcome then being broadcast to the other parties. 

\subsection{A gravitational SHIFT measurement}

The conclusion of \cite{kunjwal23a} suggests that successfully discriminating the SHIFT basis could serve as an operational signature of either the noncausal nature of a general-relativistic spacetime or an intrinsically non-classical notion of spacetime (in which ``the communication between the laboratories would still be classical, but the physical conditions for achieving this communication would be outside the realm of possibilities afforded by general-relativistic spacetimes'').

Our results refines this interpretation: rather than serving as a direct witness of noncausality (except in the case of a LOPF scenario), the ability to discriminate the SHIFT basis instead certifies the causal nonseparability of the measurement performing the discrimination. However, our findings also complement the distinction made in \cite{kunjwal23a} between a general-relativistic and a non-classical spacetime realization of the SHIFT measurement. While the former could emerge in cyclic structures such as closed timelike curves (CTCs), the latter might be realized through a gravitational quantum switch \cite{zych, paunkovic20, moller21, moller24}. Additionally, it is worth noting that both the photonic SHIFT measurement and the time-delocalized analysis (because of the SWAP operation) rely on encoding classical variables within a quantum system. While implementing a quantum control of genuinely classical variables in an photonic setting remains unclear, the concept of local operations with superposition of classical communications (LOSupCC) appears more intuitive and meaningful in a gravitational framework.

\end{document}